\documentclass[epjST]{svjour}

\usepackage[utf8]{inputenc} % Allows Unicode input
\usepackage[T1]{fontenc}    % Ensures proper font encoding

\usepackage{graphicx}
\usepackage{amsmath}
\usepackage{amssymb}
\usepackage{ulem}
\usepackage{extarrows}
\usepackage{xspace}

\newcommand{\df}{{\rm d}}
\newcommand{\as}{a_s}
\newcommand{\ncfodyzh}{$\sigma_{\text{DY,ZH}}^{\text{NNLO}}$}
\newcommand{\nclldyzh}{$\sigma_{\text{DY,ZH}}^{\text{NNLO+NNLL}}$}
\newcommand{\ncllbdyzh}{$\sigma_{\text{DY,ZH}}^{\text{NNLO+$\overline{\rm NNLL}$}}$}
\newcommand{\ncfodywh}{$\sigma_{\text{DY,WH}}^{\text{NNLO}}$}
\newcommand{\nclldywh}{$\sigma_{\text{DY,WH}}^{\text{NNLO+NNLL}}$}
\newcommand{\ncllbdywh}{$\sigma_{\text{DY,WH}}^{\text{NNLO+$\overline{\rm NNLL}$}}$}

\usepackage{xcolor}
\title{Next to Soft Threshold Resummation for $VH$ Production}

\author{
    Arunima Bhattacharya\inst{1}
    Chinmoy Dey\inst{2} 
    M.\ C.\ Kumar\inst{2} \and
    Vaibhav Pandey\inst{2} % Add more authors as needed
}
\institute{
    Theory Division, IFIC, University of Valencia-CSIC, E-46980 Paterna, Valencia, Spain \and
    Department of Physics, Indian Institute of Technology Guwahati, Guwahati-781039,\\ Assam, India
}
\date{\today}

\abstract{We study the threshold effects for the associated production of a Higgs boson with a massive vector boson $(V=Z,W)$ in the $q\bar{q} \rightarrow V^\star \rightarrow VH$ process at the LHC. By leveraging the universality of threshold logarithms and employing soft-virtual (SV) and next-to-soft virtual (NSV) resummation techniques, we compute threshold corrections to next-to-next-to-leading logarithmic accuracy.
After matching the resummed predictions to the Next-to-Next-to-Leading order (NNLO) fixed order results, we present the invariant mass distribution to NNLO$+\overline{\text{NNLL}}$ accuracy in QCD for the current LHC energies and the total production cross sections.
The $VH$ production channel is crucial for studying the couplings of the Higgs boson to the vector bosons $(W,Z)$ and understanding the mechanism of electroweak symmetry breaking. Precision measurements of this process help test the validity of the standard model (SM) and can reveal potential deviations indicating new physics.}

\begin{document}

\maketitle

% Main text of the paper
\section{Introduction}
\label{sec:intro}

%%%%%%%%%%%%%%%%%%%%%%%%%%%%%%%%%%%%%%%%%%%%%%%%%%%%%%%%%%%%%%%%

Since its inception, the SM has satisfactorily explained many natural phenomena and made accurate predictions \cite{ATLAS:2022vkf,CMS:2022dwd}. It also explains how fundamental particles gain mass by interacting with the Higgs field. Hence, the discovery of the Higgs boson on July $4$, $2012$, by the ATLAS and CMS collaborations at the Large Hadron Collider (LHC) \cite{ATLAS:2012yve,CMS:2012qbp} was a milestone achievement. Since then, the focus has shifted towards precise measurements of the Higgs boson’s properties and interactions to test the robustness of the SM and probe potential signs of physics beyond the SM (BSM). However, the known SM falls short of explaining several observed phenomena, such as the baryon asymmetry of the universe, the nature of dark matter, or the tiny nonzero masses of neutrinos.
%
%, confirming the Standard Model’s (SM) mechanism of electroweak symmetry breaking (EWSB) \cite{Englert:1964et,Higgs:1964pj,Guralnik:1964eu}. 
Determining the CP properties of the Higgs boson is one of the many interests of various precision studies conducted worldwide. While current measurements indicate that it is a scalar with even parity \cite{CMS:2017len,ATLAS:2017azn}, efforts continue scrutinising possible deviations. Additionally, numerous BSM theories predict new physics signatures that could subtly affect Higgs production and decay. Any discrepancies in measured cross sections or kinematic distributions for Higgs processes could serve as indirect evidence for such new physics \cite{Englert:2013vua,Hespel:2015zea,Goncalves:2015mfa,Bhattacharya:2019oun,Bhattacharya:2021hae,Baglio:2023euv,Iguro:2022fel,Ahmed:2021hrf,Das:2022zie,Banerjee:2024zho}.

In this context, the associated production of the Higgs boson with a massive vector boson ($VH$, where $V = Z, W^\pm$) is an important channel for probing Higgs boson interactions and testing the SM’s electroweak (EW) sector with high precision. The $VH$ process is particularly valuable due to its sensitivity to the Higgs–vector boson coupling, enabling precise tests of the $HVV$ vertex. Its experimental signatures are well-defined, especially in leptonic decay channels such as $W \to \ell \nu$ and $Z \to \ell^+ \ell^-$, which provide clean final states with charged leptons and missing energy. These features facilitate event reconstruction and help suppress backgrounds, making the $VH$ process an effective probe of Higgs properties. Although its cross section is smaller than that of gluon-gluon fusion (ggF) and vector boson fusion (VBF), the $VH$ process remains a crucial avenue for precision Higgs studies \cite{ATLAS:2019yhn,ATLAS:2020jwz,CMS:2020gsy,ATLAS:2021wqh,CMS:2024ksn}. Furthermore, $VH$ production plays a pivotal role in Higgs decay analyses, particularly in the challenging $H \to b\bar{b}$ channel, and is instrumental in constraining the top-quark Yukawa coupling and its CP structure \cite{Englert:2013vua,Hespel:2015zea,Goncalves:2015mfa}. The presence of an accompanying vector boson enhances signal detection by mitigating QCD backgrounds, making this process a key focus of experimental efforts at ATLAS and CMS \cite{ATLAS:2019yhn,ATLAS:2020jwz,ATLAS:2020fcp,CMS:2020gsy,ATLAS:2021wqh,CMS:2024ksn,CMS:2024srp,CMS:2024tdk,CMS:2024fkb,CMS:2023vzh,ATLAS:2024yzu}. Complementing these experimental efforts, accurate theoretical predictions are essential for robust interpretations of Higgs measurements.

An important contribution to $VH$ production arises from the gluon fusion channel \cite{Dicus:1988yh,Kniehl:2011aa,Harlander:2014wda}, which, despite being a loop-induced process, provides non-negligible corrections. The leading-order (LO) $ZH$ production subprocess has been computed with full top-quark mass effects in \cite{Dicus:1988yh,Kniehl:2011aa,Kniehl:1990iva,Kniehl:1990zu}, showing a $\sim7\%$ contribution relative to the next-to-leading order (NLO) Drell-Yan (DY)-type process, albeit with a substantial scale uncertainty of about $25\%$. In contrast, the next-to-next-to-leading order (N$^2$LO) QCD DY-type correction contributes only about $3\%$ relative to NLO. The effects of soft gluons in the $gg \to ZH$ subprocess have been studied for the total cross section at next-to-leading logarithmic (NLL) accuracy \cite{Harlander:2014wda} and matched to NLO QCD in the effective field theory (EFT) approximation, yielding a correction of about $15\%$. Recently, invariant mass distributions for soft gluon resummation in gluon fusion $ZH$ production have been presented in \cite{Das:2025wbj} at similar accuracy, further extended to next-to-soft effects.

Beyond fixed-order (FO) calculations, resummation techniques play a critical role in improving theoretical precision by systematically summing large threshold logarithms that arise near the partonic threshold.
The resummation of these large threshold logarithms, specifically the soft-virtual (SV) corrections, is well-established in literature \cite{Sterman:1986aj,Catani:1989ne,Catani:1990rp,Kidonakis:1997gm,Kidonakis:2003tx,Moch:2005ba,Laenen:2005uz,Kidonakis:2005kz,Ravindran:2005vv,Ravindran:2006cg,Idilbi:2006dg,Becher:2006mr,Ahmed:2020nci} and has been widely applied to various colorless processes \cite{Catani:2003zt,Moch:2005ky,deFlorian:2007sr,Kidonakis:2007ww,Catani:2014uta,Bonvini:2014joa,Ahmed:2015sna,Schmidt:2015cea,Ahmed:2016otz,Bonvini:2016frm,Kidonakis:2017dmh,AH:2019phz,Das:2019btv,Das:2019bxi,Das:2020gie,Das:2020pzo,AH:2020cok,AH:2022elh,Das:2024auk,Banerjee:2024xdh}. These studies have shown improved predictions for inclusive cross sections and invariant mass distributions.
For example, threshold resummation has been performed to next-to-next-to-next-to-leading logarithmic (N$^3$LL) accuracy for $ZH$ production \textit{via} the DY-type channel and matched to N$^3$LO QCD FO results \cite{Das:2022zie}. These studies demonstrate improved perturbative convergence and reduced scale uncertainties, offering more reliable predictions for the invariant mass distribution of the $ZH$ pair. Specifically, for the $13.6$ TeV case, scale uncertainties decrease significantly from $4.06\%$ at LO to $0.33\%$ at N$^3$LO, and from $4.44\%$ at LO+LL to $0.58\%$ at N$^3$LO+N$^3$LL. 
At this order the EW corrections are also important, the NLO EW correction for massive gauge bosons have been performed in 
\cite{Dittmaier:2001ay,Baur:2001ze,Baur:2004ig,Arbuzov:2005dd,CarloniCalame:2006zq,Zykunov:2005tc,CarloniCalame:2007cd,Arbuzov:2007db,Dittmaier:2009cr} 
, whereas the mixed EW correction are calculated in \cite{Bonciani:2020tvf,Bonciani:2021zzf,Armadillo:2022bgm} and amount to -$1.5\%$ of NLO QCD in high invariant mass region.
These findings emphasise the necessity of incorporating all relevant production channels to achieve precise theoretical predictions.

More recently, efforts have been made to incorporate next-to-soft (NSV) threshold effects as well \cite{Kidonakis:1996aq,Moch:2009hr,Soar:2009yh,Bonocore:2015esa,DelDuca:2017twk,Beneke:2018gvs,Bahjat-Abbas:2019fqa,Beneke:2019oqx,Beneke:2019mua,Moult:2019mog,Liu:2020tzd,vanBeekveld:2019cks,Das:2020adl,AH:2020iki,AH:2021vdc,vanBeekveld:2021hhv,AH:2022lpp,Beneke:2022obx,Liu:2022ajh,Sterman:2022lki,Pal:2023vec,Das:2024pac}. 
Building upon these advancements, this work focuses on improving the predictions from \cite{Das:2022zie} by incorporating next-to-soft virtual (NSV) threshold resummed corrections to next-to-next-to-leading logarithmic ($\overline{\text{NNLL}}$) accuracy for the process $q \bar{q} \to V^\star \to VH$. By matching these results to NNLO FO calculations, we obtain high-precision predictions for the invariant mass distribution and total cross sections at LHC energies. These refined predictions provide essential theoretical support for Higgs precision physics, helping to identify potential deviations from the SM and offering valuable insights into possible new physics scenarios.

%%%%%%%%%%%%%%%%%%%%%%%%%%%%%%%%%%%%%%%%

%\section{Methods}
%\label{sec:methods}
%Describe your methods and approaches in detail.

\section{Theoretical Framework}
\label{subsec:theory}

The hadronic cross section for colorless production at the hadron collider is given by,
\begin{align}\label{eq:had-xsect}
	% {d \sigma \over d Q}\left(\tau,q^2\right)
	\sigma(Q^2)
	&=\sum_{a,b={q,\overline q,g}} 
	\int_0^1 \df x_1
	\int_0^1 \df x_2~ f_a(x_1,\mu_{f}^2) ~
	f_b(x_2,\mu_{f}^2) 
%	\nn &
%	\times
	 \int_0^1 \df z \,\,
	% 2 \hat s ~{d \hat \sigma^{ab}_I \over d q^2}\left(z,q^2\right)
	\hat{\sigma}_{ab}(z,Q^2,\mu_{f}^2)
	\delta(\tau-z x_1 x_2)\,,
\end{align}
where $\sigma(Q^2) \equiv Q^2\df \sigma/\df Q^2$ for the DY-type processes. %and $\sigma(Q^2)\equiv\sigma(M_H^2)$ for $b\bar{b}H$ process.
To obtain the total production cross section of $VH$, we integrate over the invariant mass $Q$ of the final $VH$ state. The hadronic and partonic threshold variables $\tau$ and $z$ are defined as
\begin{align}
\tau=\frac{Q^2}{S}, \qquad z= \frac{Q^2}{\hat{s}} \,,
\end{align}
where $S$ and $\hat{s}$ are the hadronic and partonic centre of mass energies, respectively.
$\tau$ and $z$ are thus related by $\tau = x_1 x_2 z$.
The partonic coefficient $\hat{\sigma}_{ab}$ can be further decomposed as follows,
\begin{align}\label{eq:partonic-decompose}
	\hat{\sigma}_{ab}(z,Q^2,\mu_F^2)
	=
	\sigma^{(0)}(Q^2) \Big[ 
		\Delta_{ab}^{\rm SV}\left(z, \mu_f^2 \right) 
		+ \Delta_{ab}^{\rm NSV}\left(z, \mu_f^2 \right) + \Delta_{ab}^{\rm hard}\left(z,\mu_f^2\right)
			\Big] \,.
\end{align}
where $\sigma^{(0)}$ represents the leading-order (LO) cross section, $\Delta_{ab}^{\rm SV}$ represents the soft-virtual (SV) partonic coefficient and $\Delta_{ab}^{\rm NSV}$ represents next-to soft-virtual (NSV) contributions. While $\Delta_{ab}^{\rm SV}$ captures all the singular terms in the $z \to 1$ limit, $\Delta_{ab}^{\rm NSV}$ contains contributions in the variable $z$ , and $ \Delta_{ab}^{\rm hard}$ contains all regular terms in $z$. 
The SV+NSV cross section in $z$-space is computed in $d=4+\varepsilon$ dimensions using \cite{AH:2020iki}
\begin{equation}
\Delta_{ab}^{}\left(z,q^{2},\mu_{R}^{2},\mu_{F}^{2}\right)=\mathcal{C}\exp\left\{ \varPsi_{ab}\left(z,q^{2},\mu_{R}^{2},\mu_{F}^{2},\varepsilon\right)\right\} \mid_{\varepsilon=0}
\label{eq:deltaNSV}
\end{equation}
where $\varPsi_{ab}\left(z,q^{2},\mu_{R}^{2},\mu_{F}^{2},\varepsilon\right)$ is a finite distribution and $\mathcal{C}$ represents convolution. 
In order to study the all-order behaviour of the coefficient function $\Delta_{cc}$, we move to Mellin (N-moment) space. In this $N$--moment space, it is convenient to use the following form of the partonic coefficient function for diagonal channel partonic processes: 
\begin{eqnarray}
\label{DeltaN}
\Delta_{cc,N}(q^2,\mu_R^2,\mu_F^2) = C_0(q^2,\mu_R^2,\mu_F^2) \exp\left(
\Psi_N^c (q^2,\mu_F^2) 
\right)\,,
\end{eqnarray}
A detailed study of this structuring is done in \cite{AH:2020iki}. The coefficient $C_0(q^2,\mu_R^2,\mu_F^2)$ contains all the process-dependent information and is independent of the Mellin moment.  $C_0(q^2,\mu_R^2,\mu_F^2)$ can be expanded in powers of $a_s(\mu_R^2)$ as
\begin{eqnarray}
\label{C0expand}
C_0^g (q^2,\mu_R^2,\mu_F^2) = \sum_{i=0}^\infty a_s^i(\mu_R^2) C_{0i}^g (q^2,\mu_R^2,\mu_F^2)\,,
\end{eqnarray}
where the coefficients $C^g_{0i}$ for $VH$ production ($V=Z,W$) are given in \cite{Das:2022zie}. Here $a_s = \alpha_s/(4 \pi)$, where $\alpha_s$ is the strong coupling constant. 
%
%The term inside the exponential can be split into an SV and an NSV part, as shown in eqn.\ \ref{eqn:SV+NSV}. 
%
%
%In order to study the all-order behavior of the coefficient function, $\Delta_{ab}$, we transform it to the $N$-moment (Mellin) space. In this Mellin space, it is convenient to use the following form of the partonic coefficient function:   
%\begin{eqnarray}
%\label{resumz}
%\Delta_{ab}(q^2,\mu_R^2,\mu_F^2,z)= C^{ab}_0(q^2,\mu_R^2,\mu_F^2) 
%~~{\cal C} \exp \Bigg(2 \Psi^{ab}_{\cal D} (q^2,\mu_F^2,z) \Bigg)\,,
%\end{eqnarray}
%The coefficient $C_0^{ab}$ is $z$ independent and can be expanded in powers of $a_s(\mu_R^2)$ as
%\begin{eqnarray}
%\label{C0expand}
%C_0^{ab} (q^2,\mu_R^2,\mu_F^2) = \sum_{i=0}^\infty a_s^i(\mu_R^2) C_{0i}^g (q^2,\mu_R^2,\mu_F^2)\,,
%\end{eqnarray}
%\ab{\textbf{where the coefficients $C^{ab}_{0i}$ are calculated in \cite{} for the process under consideration. 
%}}
%
The term inside the exponential depends on the initial channel of the process under study and can be expressed as
\begin{eqnarray}
\Psi^{cc}_{\cal D} = \Psi_{\rm{SV},N}^{cc} + \Psi_{\rm{NSV},N}^{cc} ,
\label{eqn:SV+NSV}
\end{eqnarray}
where we can split $\Psi^{cc}_{\cal D}$ in such a way that all those terms that are functions of $\log^j(N),~j=0,1,\cdots$ are
kept in $\Psi_{{\rm SV},N}^{cc}$ and the remaining terms that are proportional to $(1/N) \log^j(N),~j=0,1,\cdots$ are contained
in $\Psi_{\rm{NSV},N}^{cc}$. These coefficients solely depend on the initial state partons, which for the process under study is $q\overline{q}$. 
Following the formalism in \cite{AH:2020iki}, the $\Psi_{{\rm SV},N}^{cc}$ and $\Psi_{{\rm NSV},N}^{cc}$ to all orders can be written as, 
%\sout{A formalism predicting the all-order structure of these $\Psi_{{\rm SV},N}^{ab}$ and $\Psi_{{\rm NSV},N}^{ab}$ coefficients are known and they acquire the below forms \cite{AH:2020iki}:}
\begin{eqnarray}
\label{PsiSVN}
	\Psi_{\rm{SV},N}^{cc} = \log(g_0^{c}(a_s(\mu_R^2))) + g_1^{c}(\omega)\log(N) + \sum_{i=0}^\infty a_s^i(\mu_R^2) g_{i+2}^{c}(\omega) \,,
\end{eqnarray}
%\sout{and the function $\Psi_{\rm{NSV},N}^{ab}$ is given by}
\begin{align}
\label{PsiNSVN}
 \Psi_{\rm{NSV},N}^{cc} = {1 \over N} \sum_{i=0}^\infty a_s^i(\mu_R^2) \bigg( 
 \bar g_{i+1}^{c}(\omega)  +  h^{c}_{i}(\omega,N) \bigg)\,.
\end{align}
The resummation constants $g_i, \overline{g}_i$ and $h_i$ are available for gluon fusion and quark-antiquark annihilation initiated channels in \cite{AH:2020iki,Catani:1989ne,Moch:2005ba,AH:2019phz}.

%
%For the resummed cross section, we do the matching as below:
%\begin{equation}
% \sigma^{(\text{matched})}=\sigma^{\text{SV+NSV}}_{\text{resum}}-\sigma^{\text{SV+NSV}}\bigg|_{(\text{FO})}+\sigma^{(\text{FO})}.
%\end{equation}
%
The resummed results have been matched with the available FO results to incorporate the hard regular contribution and, simultaneously, avoid double counting of SV (NSV) logarithms. The matching with the FO is performed using the \textit{minimal prescription} \cite{Catani:1996yz} and 
for $\overline{\rm N{\it n}LL}$ resummation it reads,
\begin{align}\label{eq:MATCHING}
	Q^2\frac{\df \sigma^{\rm N{\it n}LO+\overline{N{\it n}LL}}_{ab}}{\df Q^2}
	=&
	Q^2 \frac{\df {\sigma}^{\rm N{\it n}LO}_{ab} }{\df Q^2}
	+
	%\sum_{a,b \in \{q,\bar{q}\}}
        \sum_{ab \in \{gg, q\bar{q}\} }
        \widehat{\sigma}^{(0)}_{ab}(Q^2)
	\int_{c-i\infty}^{c+i\infty}
	\frac{\df N}{2\pi i}
	\tau^{-N}
	f_{a,N}(\mu_f)
	f_{b,N}(\mu_f) \\
	&\times 
	\Bigg( 
		Q^2\frac{\df \widehat{\sigma}_{N,ab}^{\overline{\rm N{\it n}LL}}}{\df Q^2} 
		- 	
		Q^2\frac{\df \widehat{\sigma}_{N,ab}^{\overline{\rm N{\it n}LL}}}{\df Q^2} \Bigg|_{\rm tr}
	\Bigg) \,.
\end{align}    
In the later sections, we present these matched results in the $\overline{N}$-scheme ($\overline{N} = N e^{\gamma_E}$, where $\gamma_E$ is the Euler–Mascheroni constant) following the approach given in \cite{AH:2020cok}. 
We also define K-factors for the FO $(K_{nm})$, SV resummed $(R_{nm})$ and NSV resummed $(\overline{R}_{nm})$ cross sections as below:

\begin{align}
	{ K}_{nm}
	= 
	\frac{\sigma^{\text{N}{\it n}\text{LO}}}{\sigma_c^{\text{N}{\it m}\text{LO}}} 
	,
	{R}_{nm} 
	= 
	\frac{\sigma^{\text{N}{\it n}\text{LO} + \text{N}{\it n}\text{LL}}}{\sigma_c^{\text{N}{\it m}\text{LO}}} \text{ and }
	{ \overline{R}}_{nm} 
	= 
	\frac{\sigma^{\text{N}{\it n}\text{LO} + \overline{\text{N}{\it n}\text{LL}}}}{\sigma_c^{\text{N}{\it m}\text{LO}}} \,.
	\label{eq:ratio}
\end{align}
Using these tools and methodologies, we have calculated the invariant mass-distribution, $7$-point scale uncertainties, PDF intrinsic uncertainties, scale uncertainties and the total production cross sections for $q\overline{q}\rightarrow V^{\star} \rightarrow VH$. These results are discussed in the next section of the article. 
%

% %\ab{
% \cd{The resummation of large soft (SV) logarithms is well-established in the literature \cite{Sterman:1986aj,Catani:1989ne,Catani:1990rp,Kidonakis:1997gm,Kidonakis:2003tx,Moch:2005ba,Laenen:2005uz,Kidonakis:2005kz,Ravindran:2005vv,Ravindran:2006cg,Idilbi:2006dg,Becher:2006mr,Ahmed:2020nci} 
% and has been widely applied to various colorless processes \cite{Catani:2003zt,Moch:2005ky,deFlorian:2007sr,Kidonakis:2007ww,Catani:2014uta,Bonvini:2014joa,Ahmed:2015sna,Schmidt:2015cea,Ahmed:2016otz,Bonvini:2016frm,Kidonakis:2017dmh,AH:2019phz,Das:2019btv,Das:2019bxi,Das:2020gie,Das:2020pzo,AH:2020cok,AH:2022elh,Das:2024auk,Banerjee:2024xdh} 
% leading to improved predictions for inclusive cross-sections and invariant mass distributions. More recently, efforts have been made to incorporate next-to-soft (NSV) threshold effects as well \cite{Kidonakis:1996aq,Moch:2009hr,Soar:2009yh,Bonocore:2015esa,DelDuca:2017twk,Beneke:2018gvs,Bahjat-Abbas:2019fqa,Beneke:2019oqx,Beneke:2019mua,Moult:2019mog,Liu:2020tzd,vanBeekveld:2019cks,Das:2020adl,AH:2020iki,AH:2021vdc,vanBeekveld:2021hhv,AH:2022lpp,Beneke:2022obx,Liu:2022ajh,Sterman:2022lki,Pal:2023vec}.} \textbf{COMMENT:} Added this part of \cd{} to the Introduction in the second last paragraph.}

\section{Results and Discussion}
\label{sec:results}
For the numerical evaluations we use the EW couplings in $G_{\mu}$ scheme, where  we use $G_{F}=1.1663788 \times 10^{-5}$ , $m_{Z} = 91.1880$ GeV and  $m_{W} = 80.3692$ GeV to calculate the fine structure constant $\alpha = G_F (8 \sin^2{\theta_{W}} \cos^2{\theta_{W}} m_{Z}^2 )/(4 \sqrt{2} \pi) $ with $\cos^2{\theta_{W}}=m_{W}^{2}/m_{Z}^{2}$. The decay widths are taken to be $\Gamma_Z = 2.4955$ GeV and $\Gamma_W = 2.085$ GeV. The mass of the Higgs boson is taken to be $m_H = 125.2$ GeV. 
Our calculations are done with five massless quark flavours ($n_{f}=5$). 
In our calculation, the CKM matrix elements are $V_{ud} = 0.97446$,  $V_{us} = 0.22452$,  $V_{ub} = 0.00365$,  $V_{cd} = 0.22438$, $V_{cs} = 0.97359$ and  $V_{cb} = 0.04214$. 
The partonic cross sections are folded with the \texttt{MSHT20} \cite{Bailey:2020ooq} sets of parton distribution functions (PDFs) exctracted at NNLO level, and the strong coupling constant is taken from \texttt{LHAPDF} \cite{Buckley:2014ana} with $\alpha_s(m_Z) =0.118$. The central choice of scale for the unphysical renormalization and factorization scales is taken to be the invariant mass $Q$ of the $VH$ final state ($\mu_{R}=\mu_{F}=Q$). The conventional 7-point scale variation is performed by varying the unphysical scales in the range such that $\left| \ln{(\mu_{R}/\mu_{F})} \right| \leq \text{ln}\,2$.

We utilise \texttt{n3loxs} package to compute the FO results for neutral and charged DY type $VH$ production process \cite{Anastasiou:2018qmv,Baglio:2022wzu}.
%Specifically, we employ \texttt{n3loxs} to compute the neutral DY corrections while explicitly neglecting the contributions from the virtual photon $\gamma^{\star}$ in our calculation.
At the NNLO level for the $ZH$ production process, in addition to the quark annihilation process, there will also be contributions coming the loop induced gluon fusion channel and top-quark loops. Apart from these, there will also contribution from the bottom annihilation process where the Higgs boson directly couples to the bottom quark. We have taken all these contributions from 
% Additionally, the FO contributions from the gluon-gluon (gg) and quark-gluon (qg) channels in the invariant mass distribution study are obtained using 
\texttt{vh@nnlo} \cite{Forte:2016mus,Boughezal:2016vh}.

\begin{figure}[ht!]
\centerline{
\includegraphics[scale=0.25]{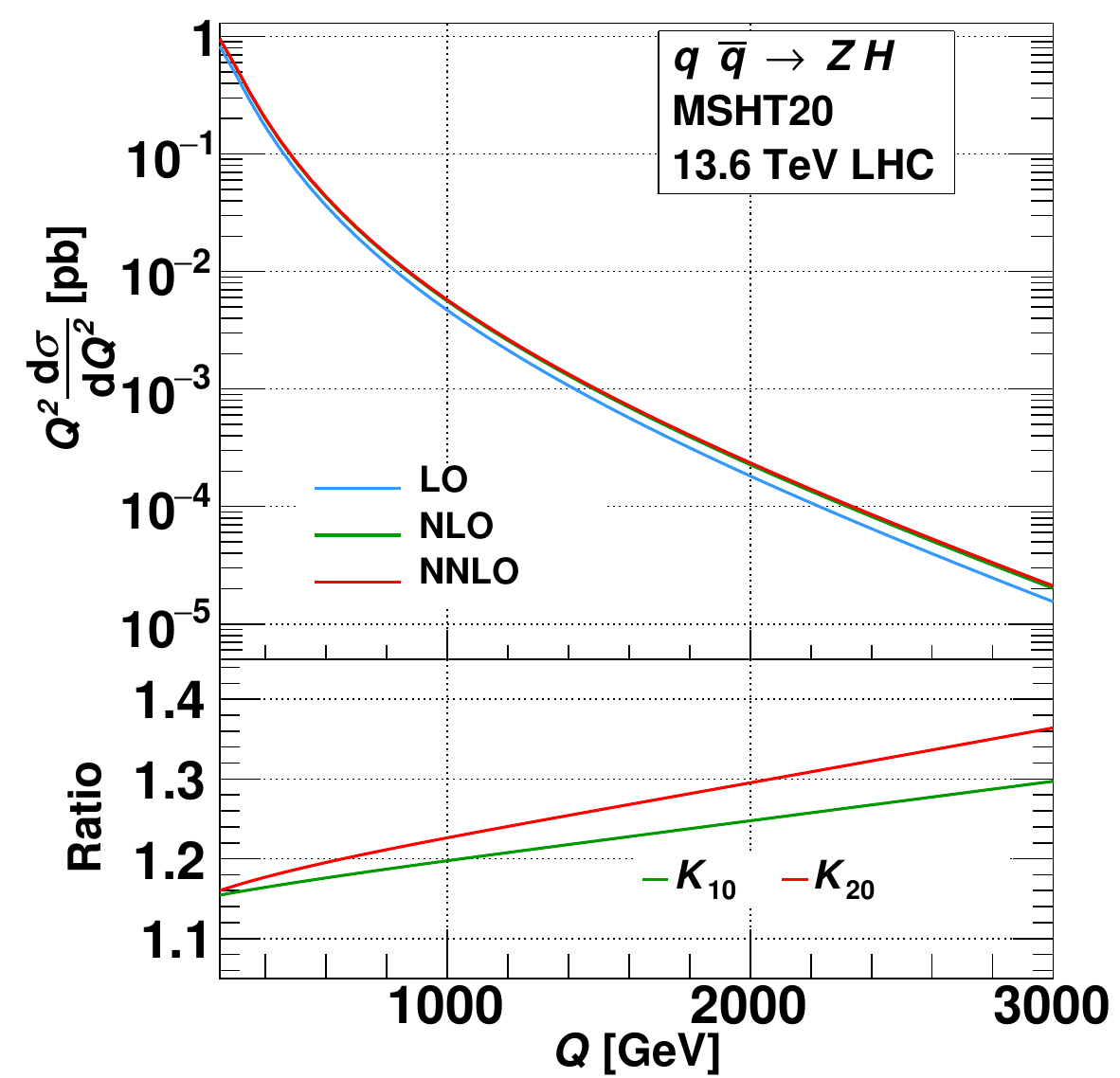}
\includegraphics[scale=0.25]{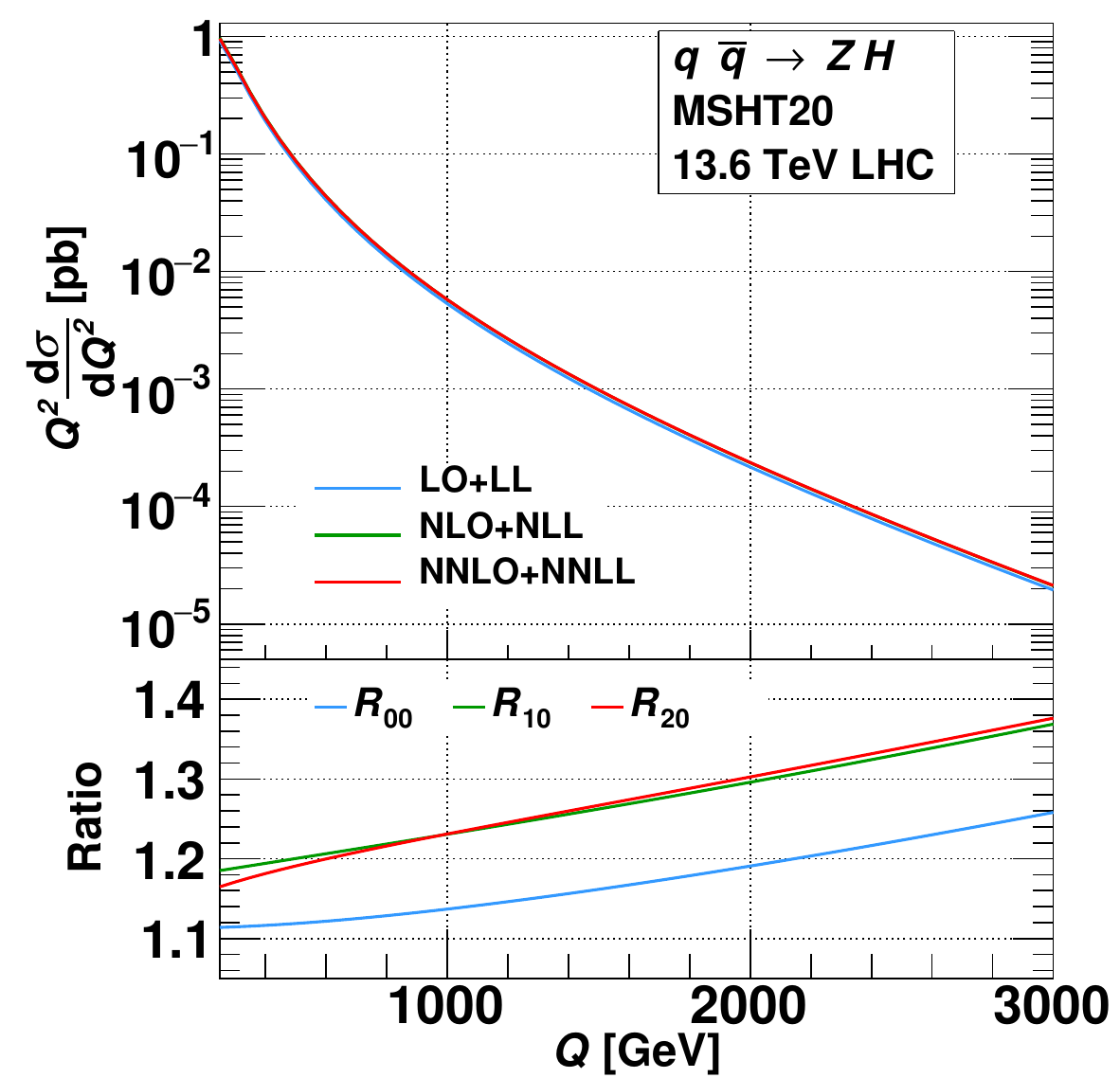}
\includegraphics[scale=0.25]{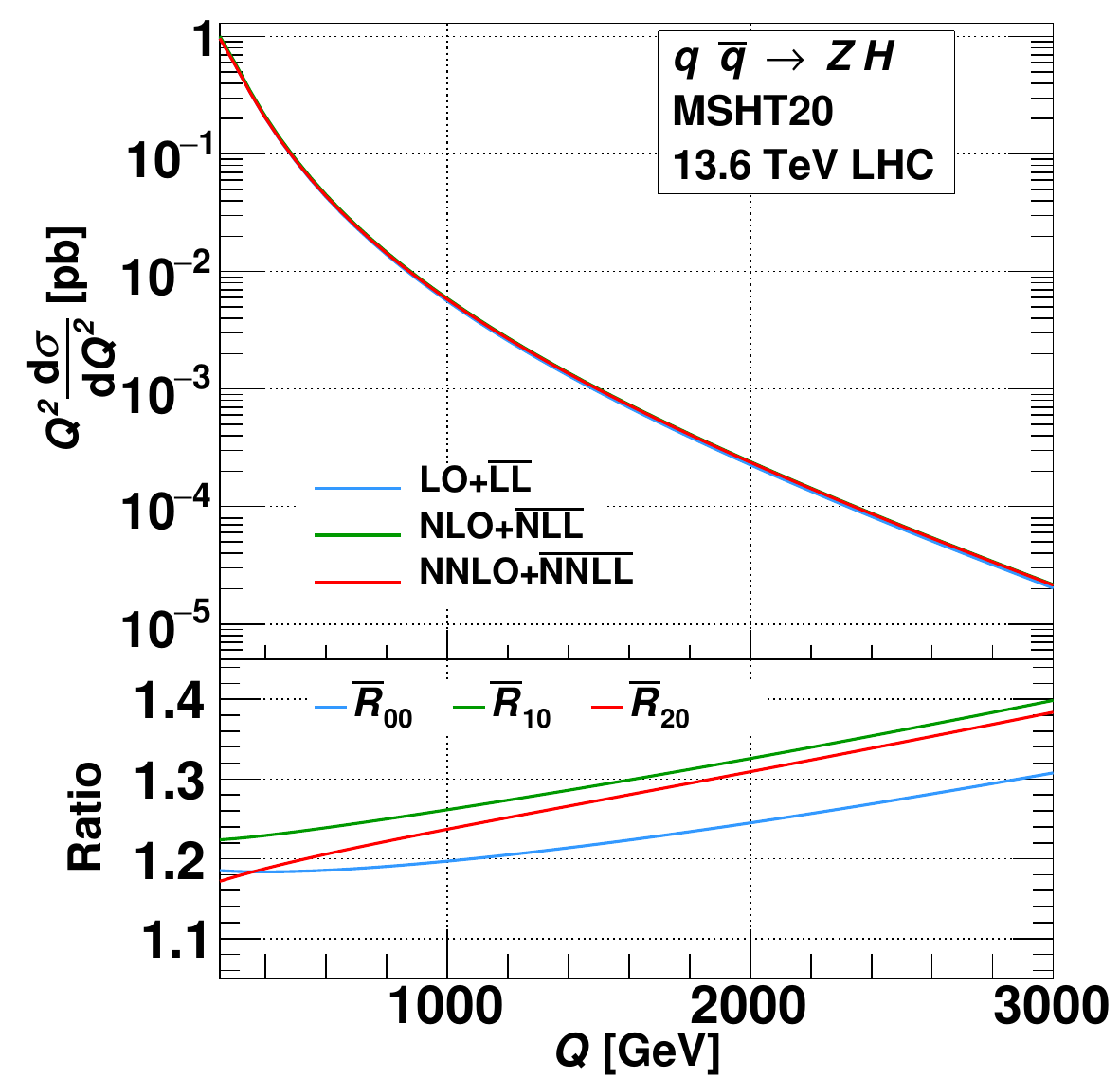}
}
\vspace{-2mm}
\caption{\small{The invariant mass distribution with K-factor FO (left), SV (middle) and NSV (right) are shown for the $q\bar{q} \to ZH$ process at $13.6$ TeV LHC. }}
\label{fig:inv_ZH}
\end{figure}

\begin{figure}[ht!]
\centerline{
\includegraphics[scale=0.25]{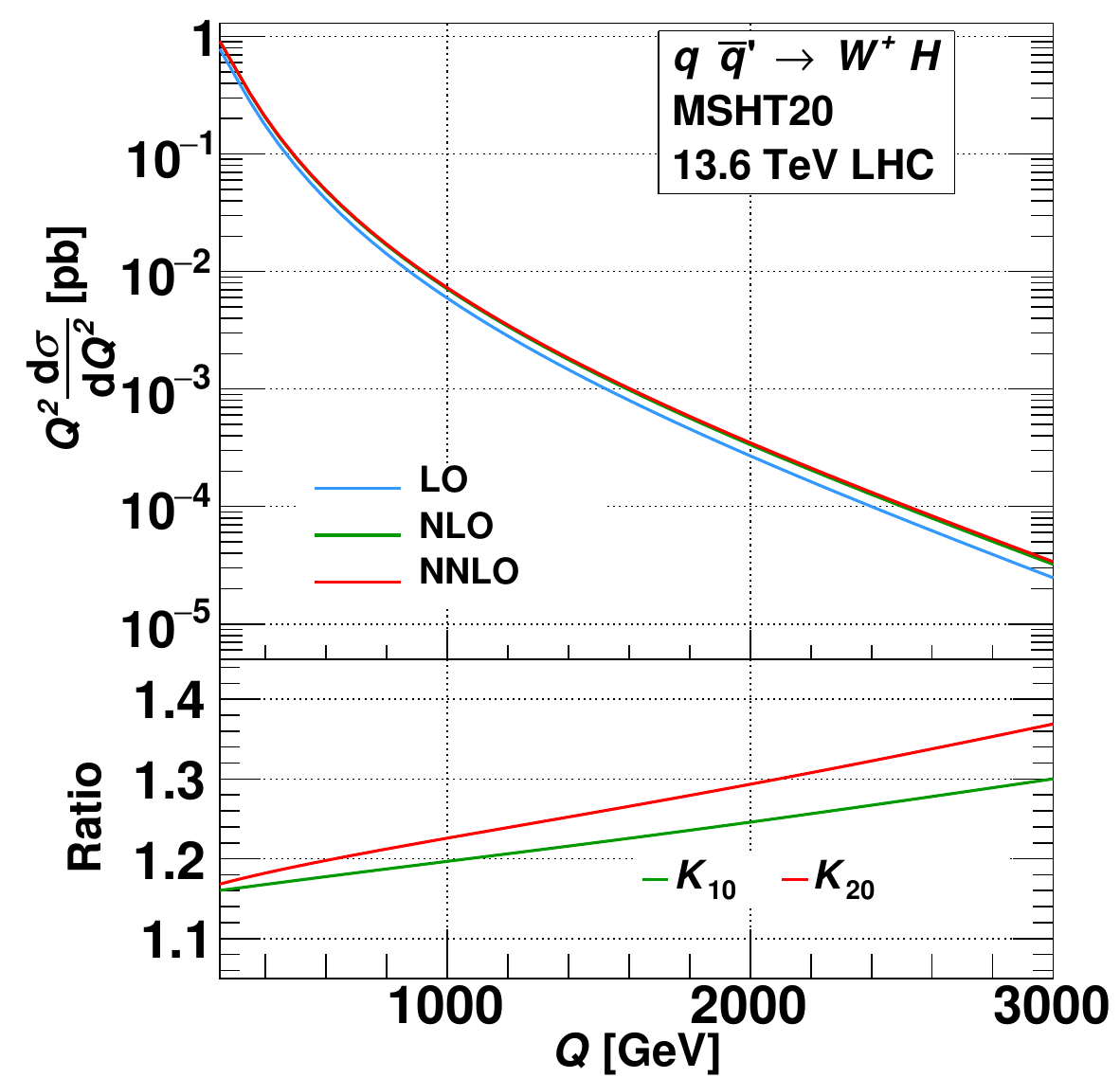}
\includegraphics[scale=0.25]{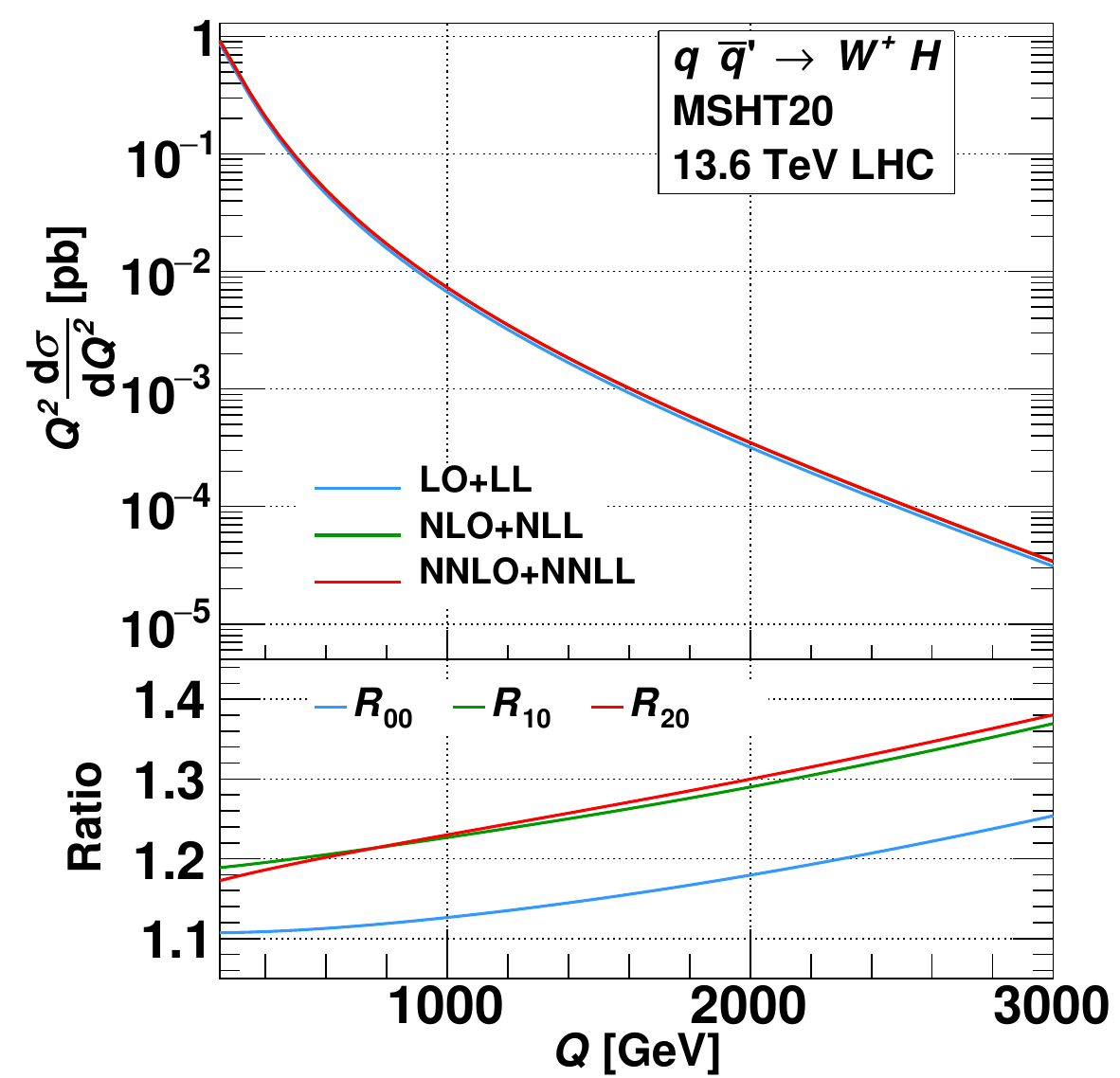}
\includegraphics[scale=0.25]{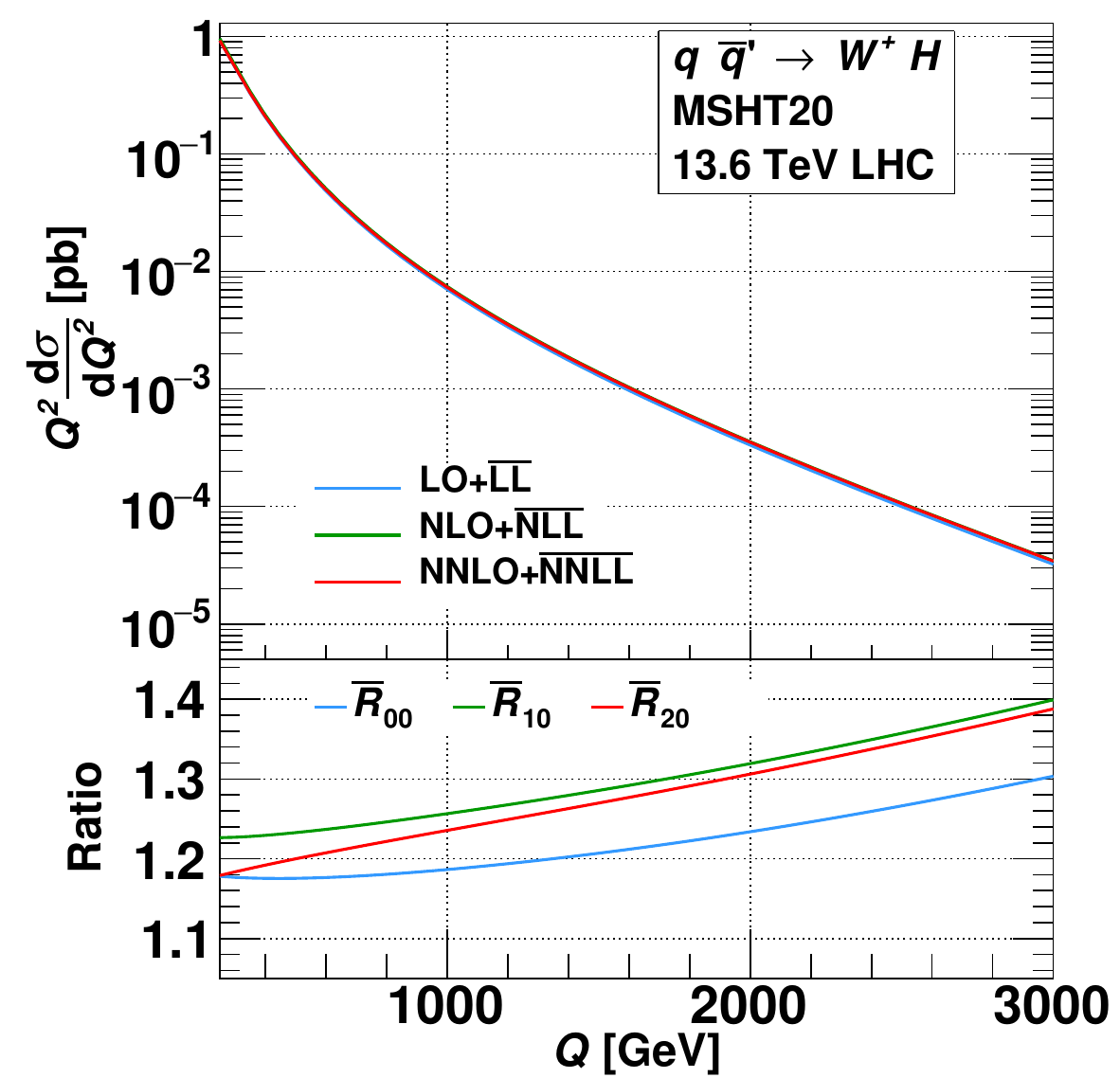}
}
\vspace{-2mm}
\caption{\small{The invariant mass distribution with K-factor FO (left), SV (middle) and NSV (right) are shown for the $q\bar{q}^{\prime} \to W^+H$ process at $13.6$ TeV LHC. }}
\label{fig:inv_WpH}
\end{figure}

\begin{figure}[ht!]
\centerline{
\includegraphics[scale=0.25]{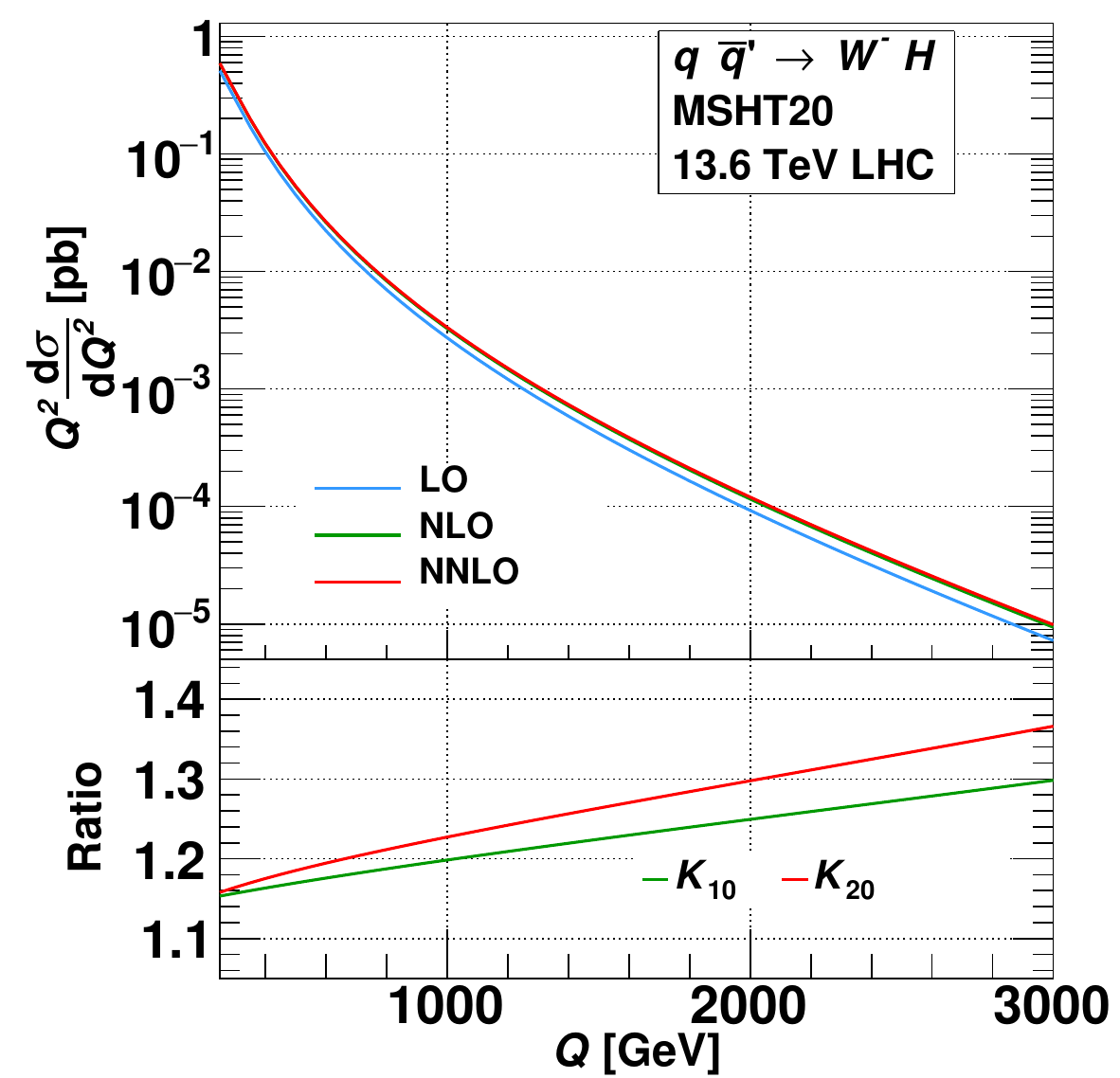}
\includegraphics[scale=0.25]{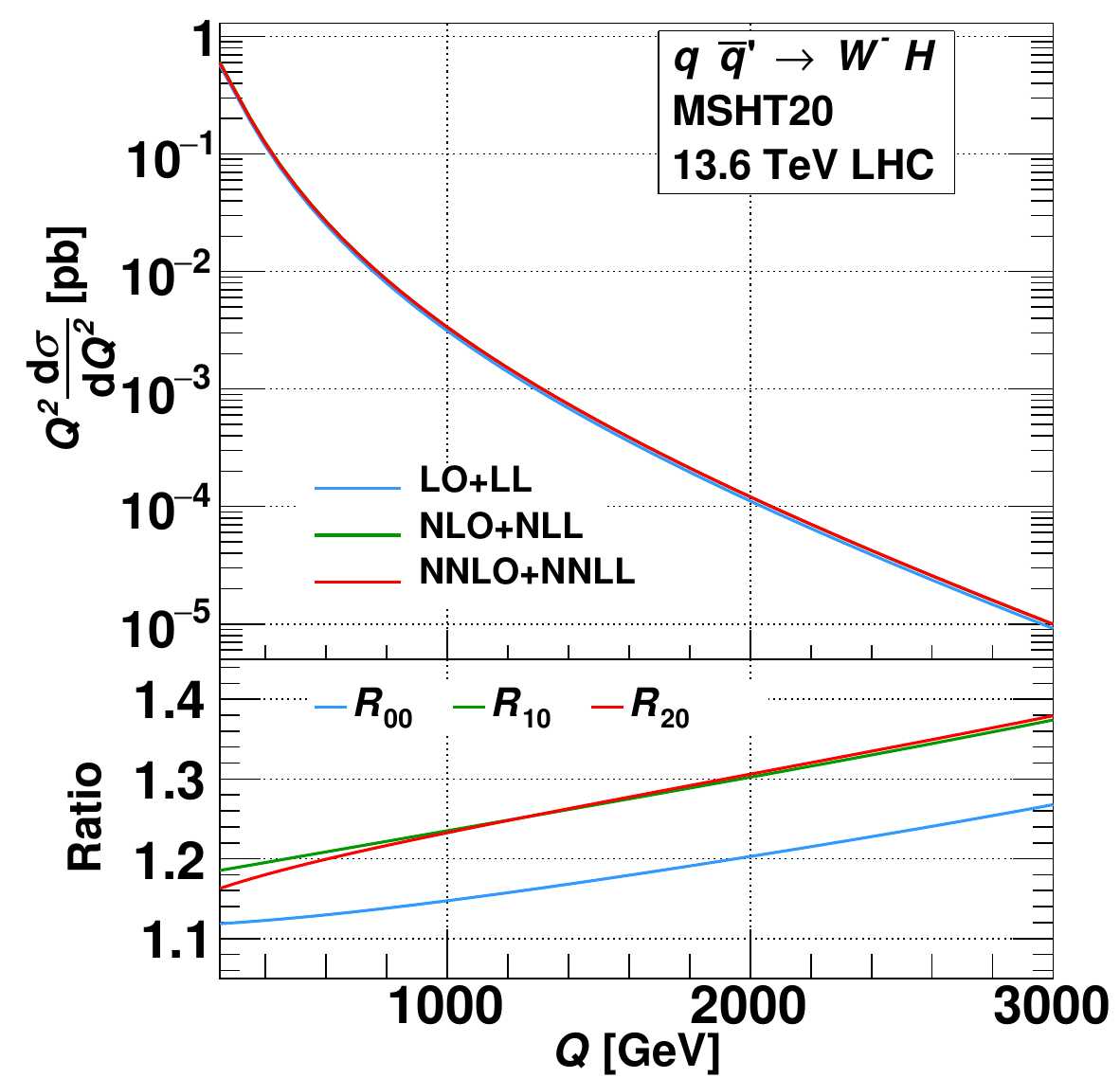}
\includegraphics[scale=0.25]{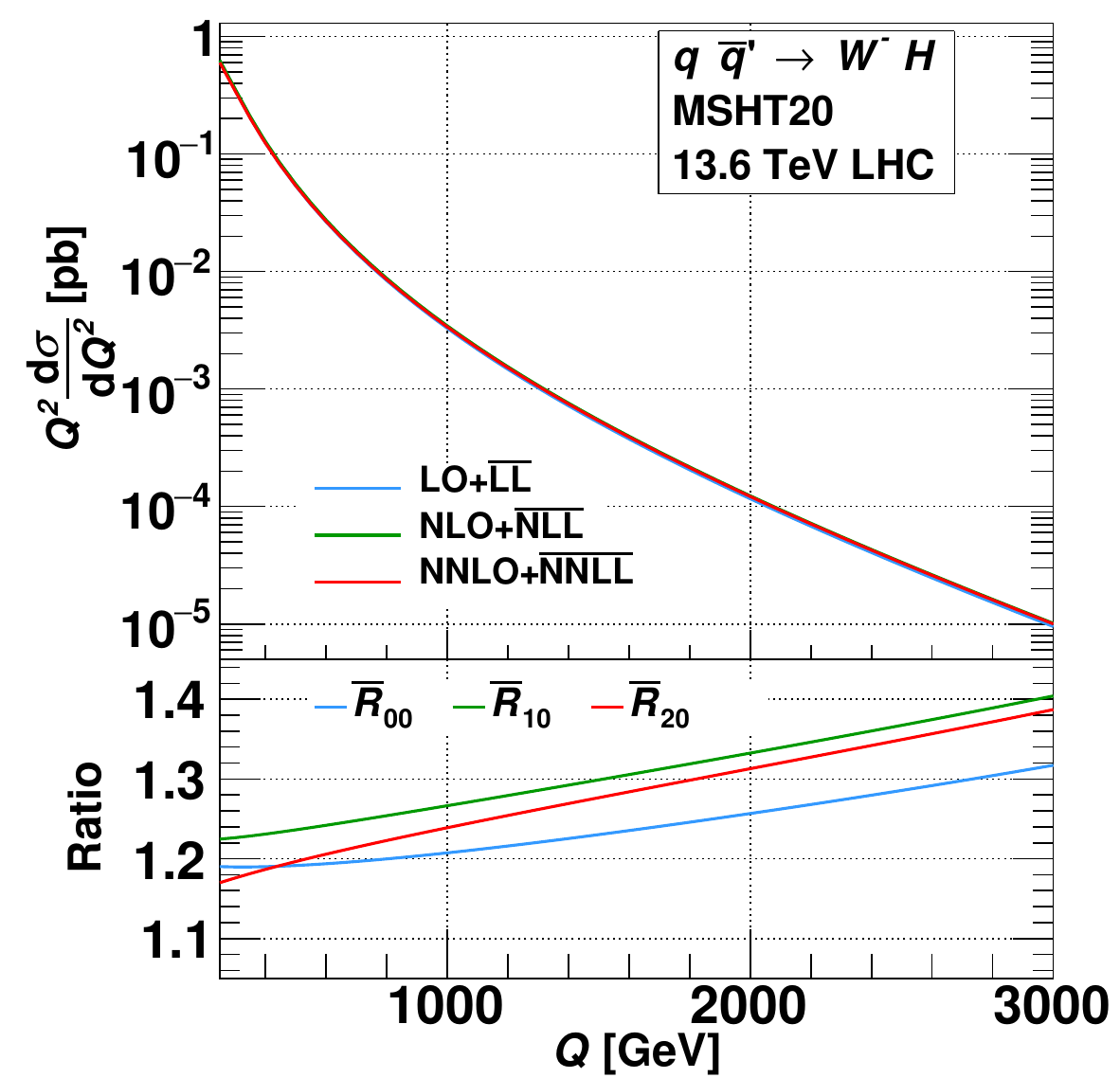}
}
\vspace{-2mm}
\caption{\small{The invariant mass distribution with K-factor FO (left), SV (middle) and NSV (right) are shown for the $q\bar{q}^{\prime} \to W^-H$ process at $13.6$ TeV LHC. }}
\label{fig:inv_WmH}
\end{figure}

In fig. \ref{fig:inv_ZH}, fig. \ref{fig:inv_WpH} and fig.  \ref{fig:inv_WmH}, we present the invariant mass distribution for the $ZH$ and  $W^{\pm}H$   production processes up to NNLO$+ \overline{\text{NNLL}}$ accuracy from $250$ GeV to $3000$ GeV at $13.6$ TeV LHC. The cross section decreases with the diminution of the $q \bar{q}$ flux from the low to the high $Q$ region. To understand the importance of these corrections better, the corresponding K-factors (defined in eq. \ref{eq:ratio}) are shown in the lower panel of the same plots. The fixed order K-factors \( K_{10} \) and \( K_{20} \) for $ZH$ production acquire values of up to $1.30$ and $1.36$, respectively, in the high $Q$ region around 3 TeV. In the same $Q$ range, the K-factors associated with the SV resummed corrections, namely \( R_{10} \) and \( R_{20} \), reach approximately $1.37$ and $1.38$. Furthermore, the inclusion of NSV logarithms increase the K-factors, \( \overline{R}_{10} \) and \( \overline{R}_{20} \), to approximately $1.40$ and $1.38$, respectively. 
A comparable pattern is observed for charged DY processes, where the NLO K-factor $(K_{10})$ and the NNLO K-factor $(K_{20})$ undergo substantial modifications upon incorporating the resummed results to NLO+\(\overline{\text{NLL}}\) (\( \overline{R}_{10} \)) and NNLO+\(\overline{\text{NNLL}}\) (\( \overline{R}_{20} \)) accuracy. 
A key observation in figs.\ \ref{fig:inv_ZH}, \ref{fig:inv_WpH} and \ref{fig:inv_WmH} is that the addition of NSV resummation to the SV resummed results at two-loop order negatively impacts the K-factor values, denoted by \( R_{20} \) and \( \overline{R}_{20} \). 
% We note here that the resummation is performed in the $\bar{N}$-scheme. However, if the resummation is performed in the $N$-scheme, the NSV resummation results are found to be positive, and such a behaviour is already seen in the literature (see fig.14 of \cite{AH:2021kvg}).  
This can be attributed to our choice of working in the $\bar{N}$ scheme. To understand this interplay between schemes, check \cite{AH:2021kvg} where the authors present the resummed predictions for inclusive cross section for DY production. In fig.\ $14$ of \cite{AH:2021kvg}, the $\text{NLO + }\overline{\text{NLL}}$ results are shown to be overestimated in the $\overline{N}$-scheme compared to the ones in the $N$-scheme. The same figure also illustrates that the results at two-loop accuracy $(\text{NNLO + }\overline{\text{NNLL}})$ in both the schemes are comparable. This indicates a decrease in values from the $\text{NLO + }\overline{\text{NLL}}$ results in the $\overline{N}$-scheme, while showing an increase in values in the $N$-scheme. It is also worth noting that the rate of convergence of the perturbation series is better for $\overline{N}$-scheme compared to $N$-scheme in Ref. \cite{AH:2020cok}.
    % 
    %
    %\ab{\textbf{COMMENT: The values of $R_{20}$ and $\overline{R}_{20}$ are same = 1.38!  Is that correct?}}
    %
%%%%%%%%%%% Fig: qq -> ZH Invariant mass
%%%%%%%%%%%%
%%%%%%%%%%% Fig: qq -> WpH Invariant mass
%%%%%%%%%%%%
%%%%%%%%%%% Fig: qq -> WmH Invariant mass
%%%%%%%%%%%%

\begin{figure}[ht!]
\centerline{
\includegraphics[scale=0.25]{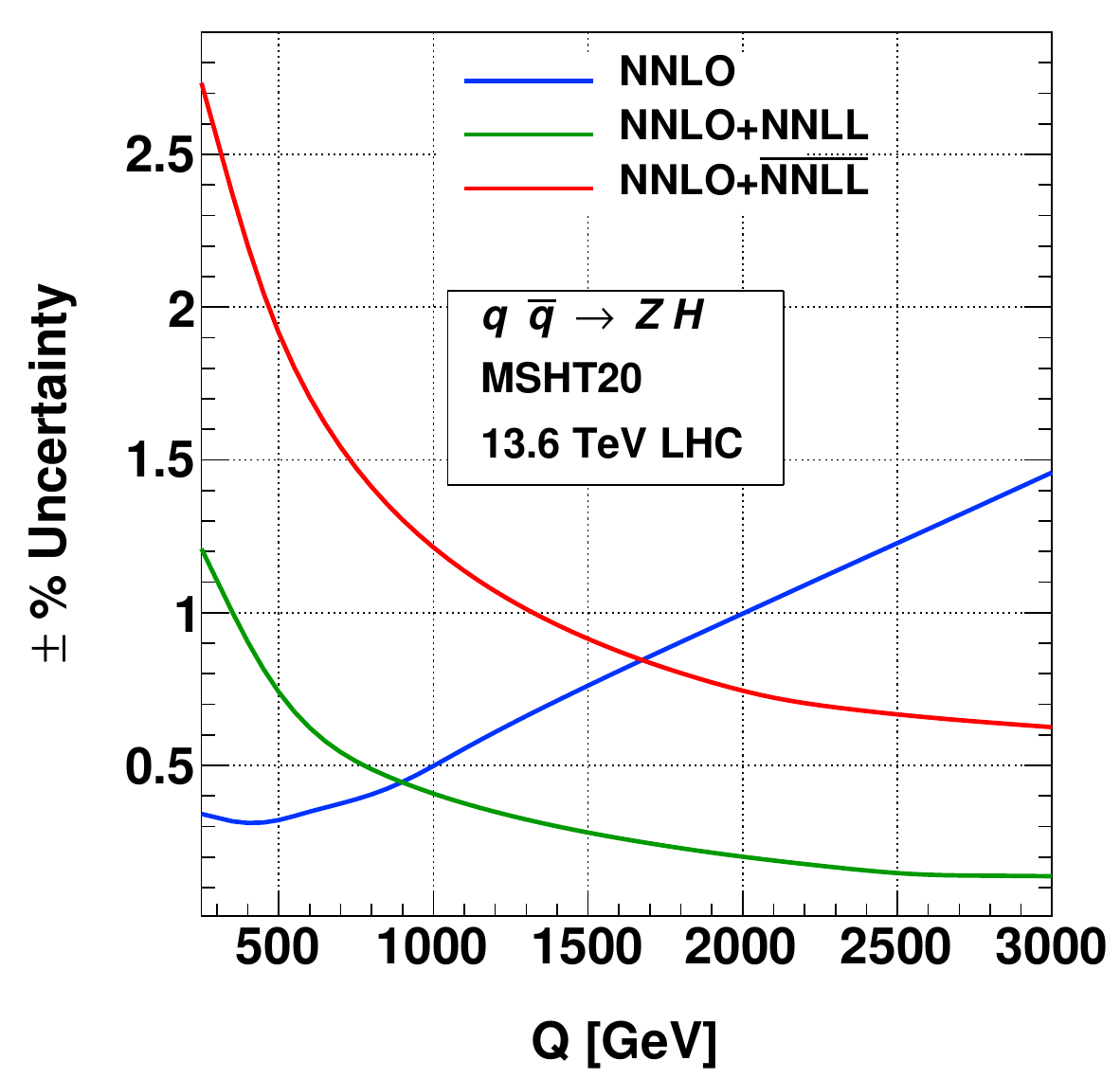}
\includegraphics[scale=0.25]{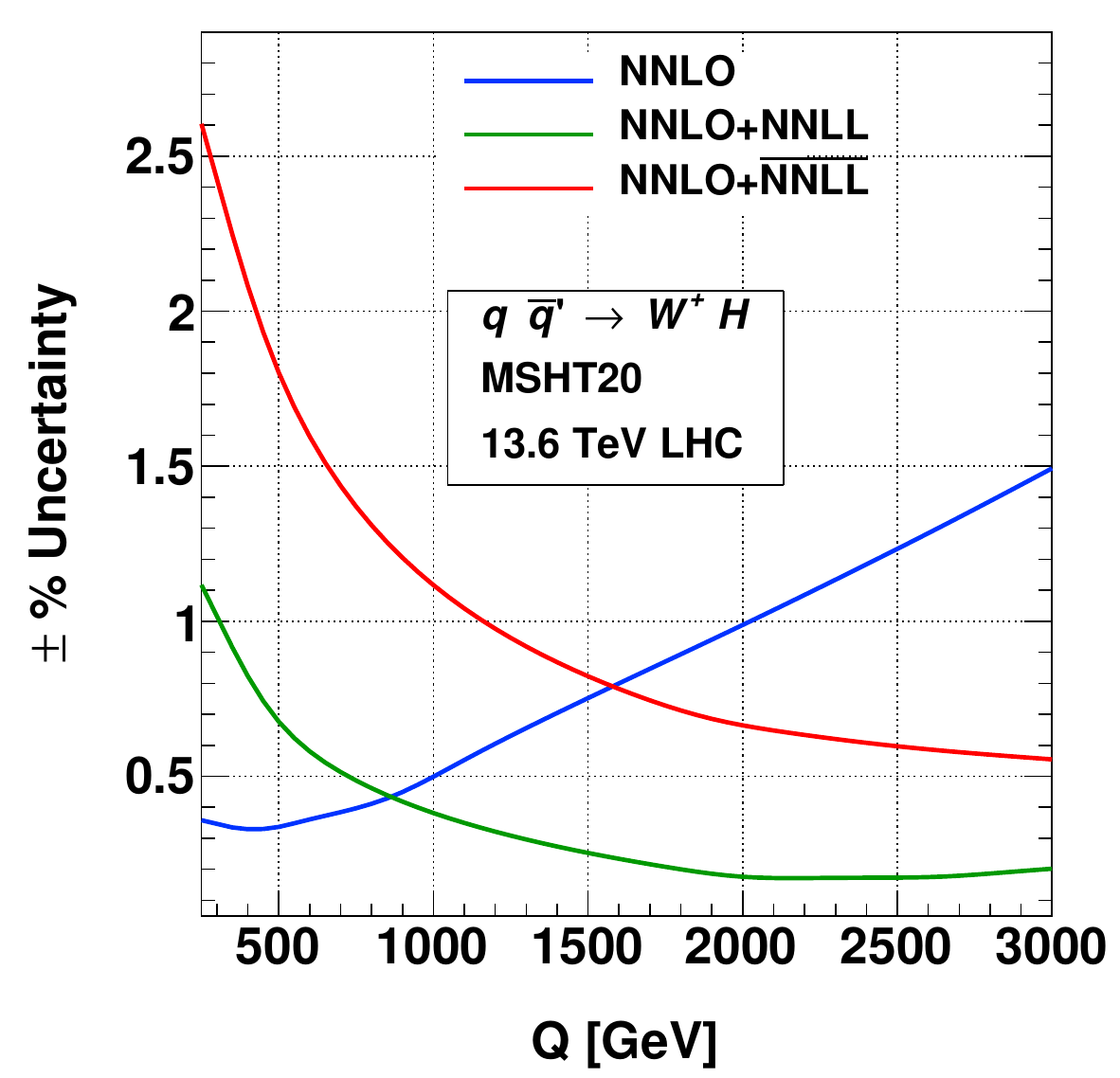}
\includegraphics[scale=0.25]{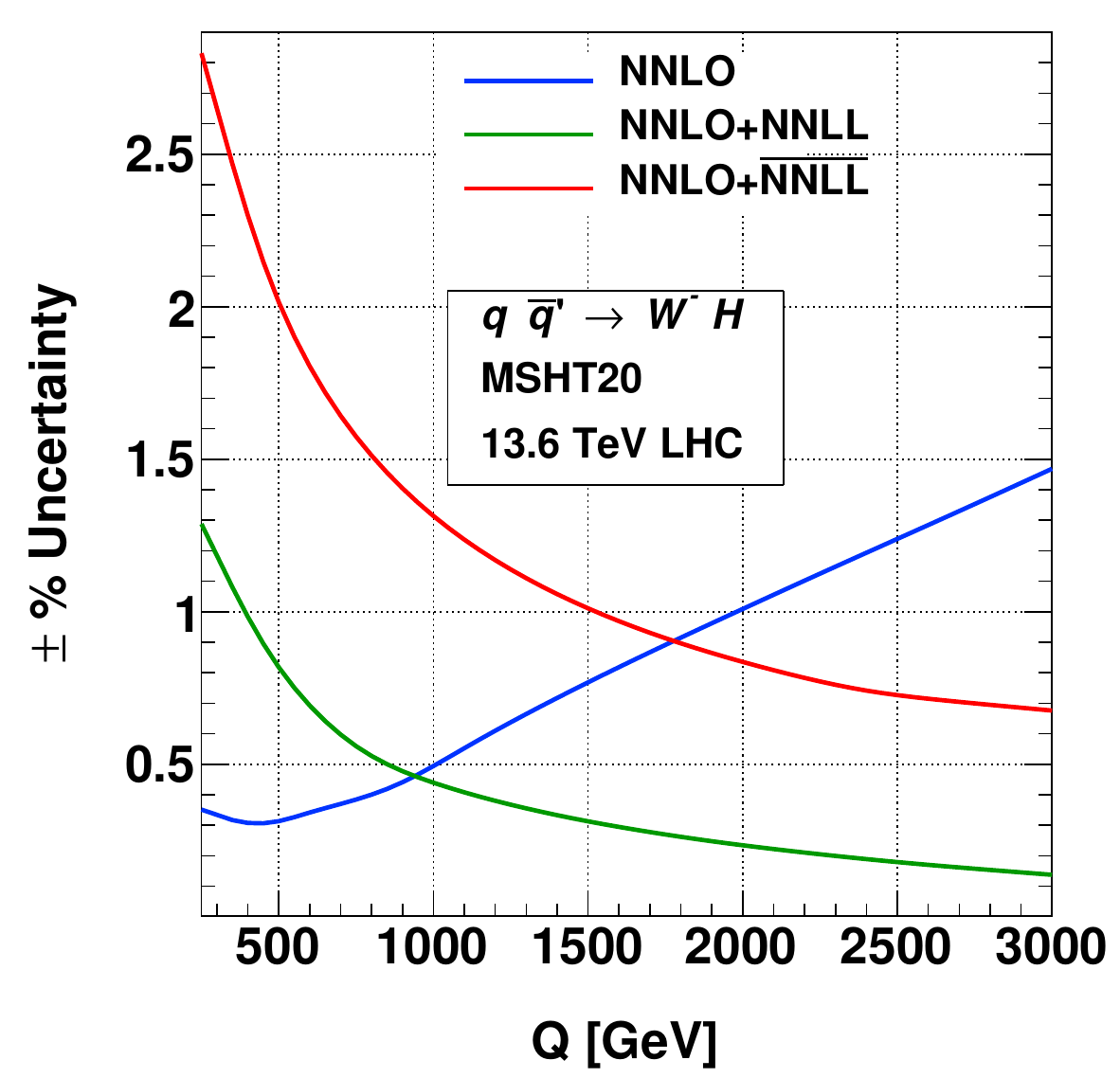}
}
\vspace{-2mm}
\caption{\small{The 7-point scale uncertainty $ZH$ (left), $W^+H$ (middle) and $W^-H$ (right) are shown at $13.6$ TeV LHC. }}
\label{fig:scale_VH}
\end{figure}

Fig. \ref{fig:scale_VH} depicts the 7-point scale uncertainties for $ZH$ and $W^{\pm}H$ production processes at NNLO, NNLO+NNLL, and NNLO+$\overline{\text{NNLL}}$ accuracy, ranging from $250$ GeV to $3000$ GeV at the $13.6$ TeV LHC. In the low $Q$ region $(<900 \text{ GeV})$, the uncertainties increase as we move from NNLO to NNLO+NNLL and then to NNLO+$\overline{\text{NNLL}}$. On the other hand, in the high $Q$ region, the SV resummed corrections reduce the scale uncertainties compared to the FO results. Conversely, the NSV resummed corrections demonstrate higher uncertainties than NNLO and NNLO+NNLL. 
To understand this behaviour, we need to recollect that for $VH$ production, apart from the $q\bar{q}$ initiated sub-process, we have contributions from other partonic channels like $qg$ and $gg$.
%    \begin{equation}
%        \sigma_{pp \rightarrow VH}^{\text{NNLO}} = \sigma_{q \bar{q}\rightarrow VH}^{\text{NNLO}} + \sigma_{q g\rightarrow VH}^{\text{NNLO}} + \sigma_{g g\rightarrow VH}^{\text{NNLO}}
%    \end{equation} 
However, in this work, we are focusing on the threshold resummation for the ${q \bar{q}}$ channel only. The other parton channels (e.g. $qg$) will not contribute to the threshold (SV) logarithms. However, such channels do contribute to the NSV logarithms and are not considered in the present work.
%effects of threshold logarithms from the other partonic channels are not in the scope of this work. 
    %
Therefore, we observe that incorporating the NSV resummation leads to an increase in scale uncertainties. 
    %
    %\textbf{HOW ARE THE PLOTS 4, 5 AND 6 DIFFERENT FROM PLOT 7? STATE CLEARLY! THE MEANING OF THE SUBSCRIPT $q \bar{q}$ in FIGURE 7.} 
    %\cd{$q\bar{q}$ means only $q\bar{q}\to ZH$ contribution at NNLO from FO and resum results are same, i.e. no contribution from $qg$ and $gg$ at NNLO. Separately, we study the $q\bar{q}$ channel NSV $\mur$ uncertainty, and it reduces the uncertainty because of NSV resummation.}
    %$q\bar{q}$ means only $q\bar{q}\to ZH$ contribution at NNLO from FO  and resum results are same, i.e. no contribution from $qg$ and $gg$ at NNLO. 
    %Separately, we study the $q\bar{q}$ channel NSV $\mur$ uncertainty, and it reduces the uncertainty because of NSV resummation.
%

\begin{figure}[ht!]
\centerline{
\includegraphics[scale=0.25]{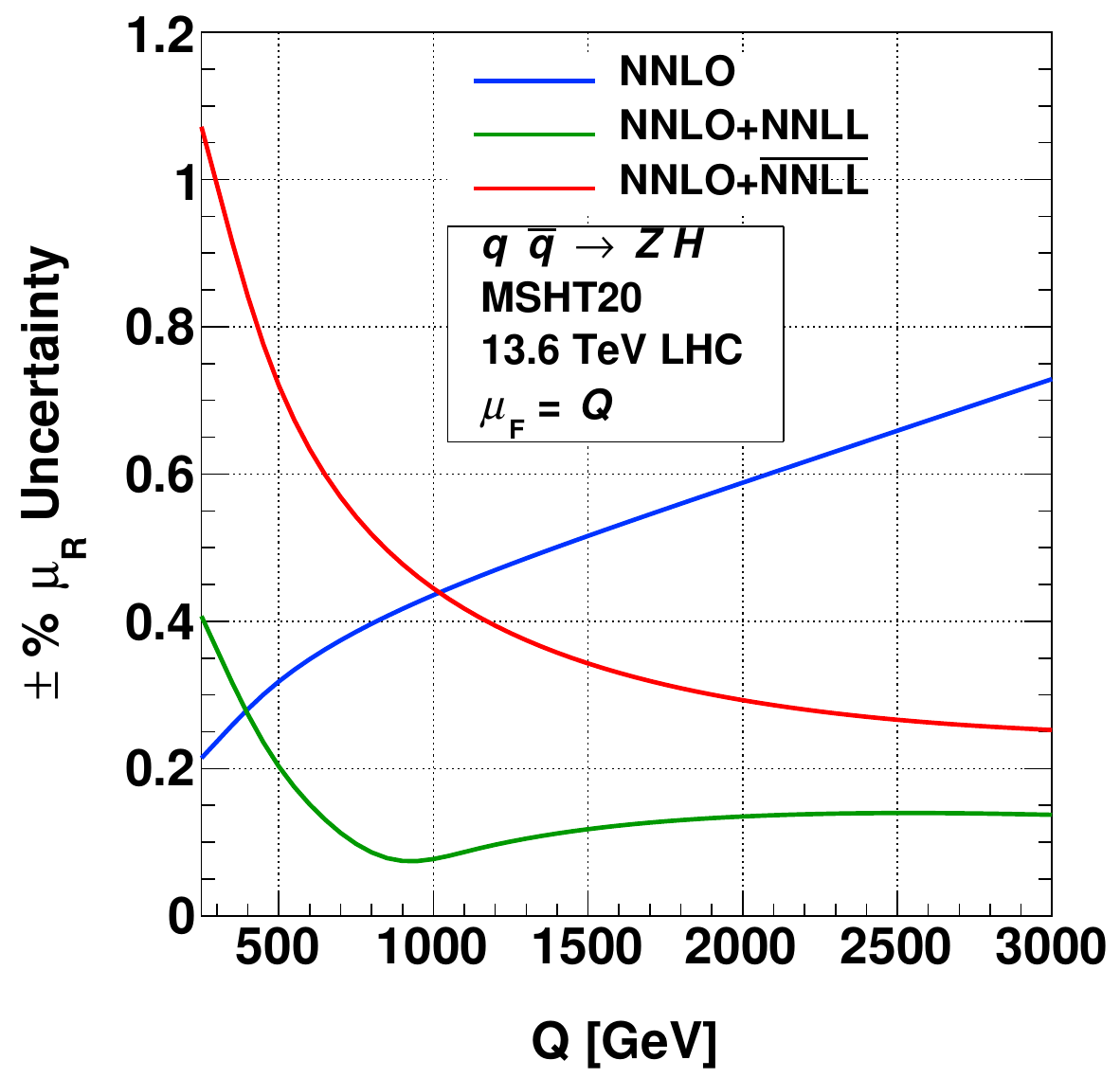}
\includegraphics[scale=0.25]{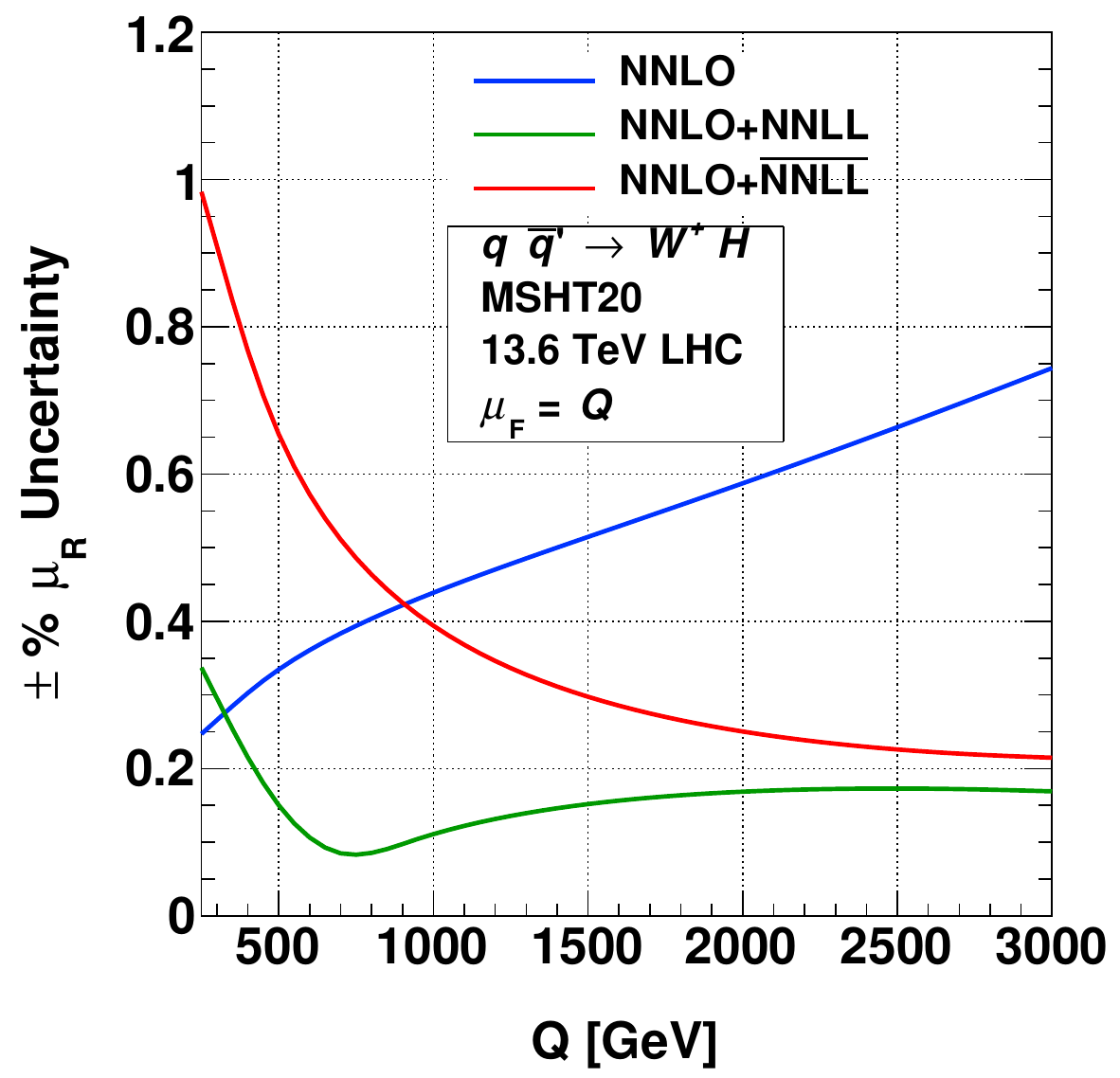}
\includegraphics[scale=0.25]{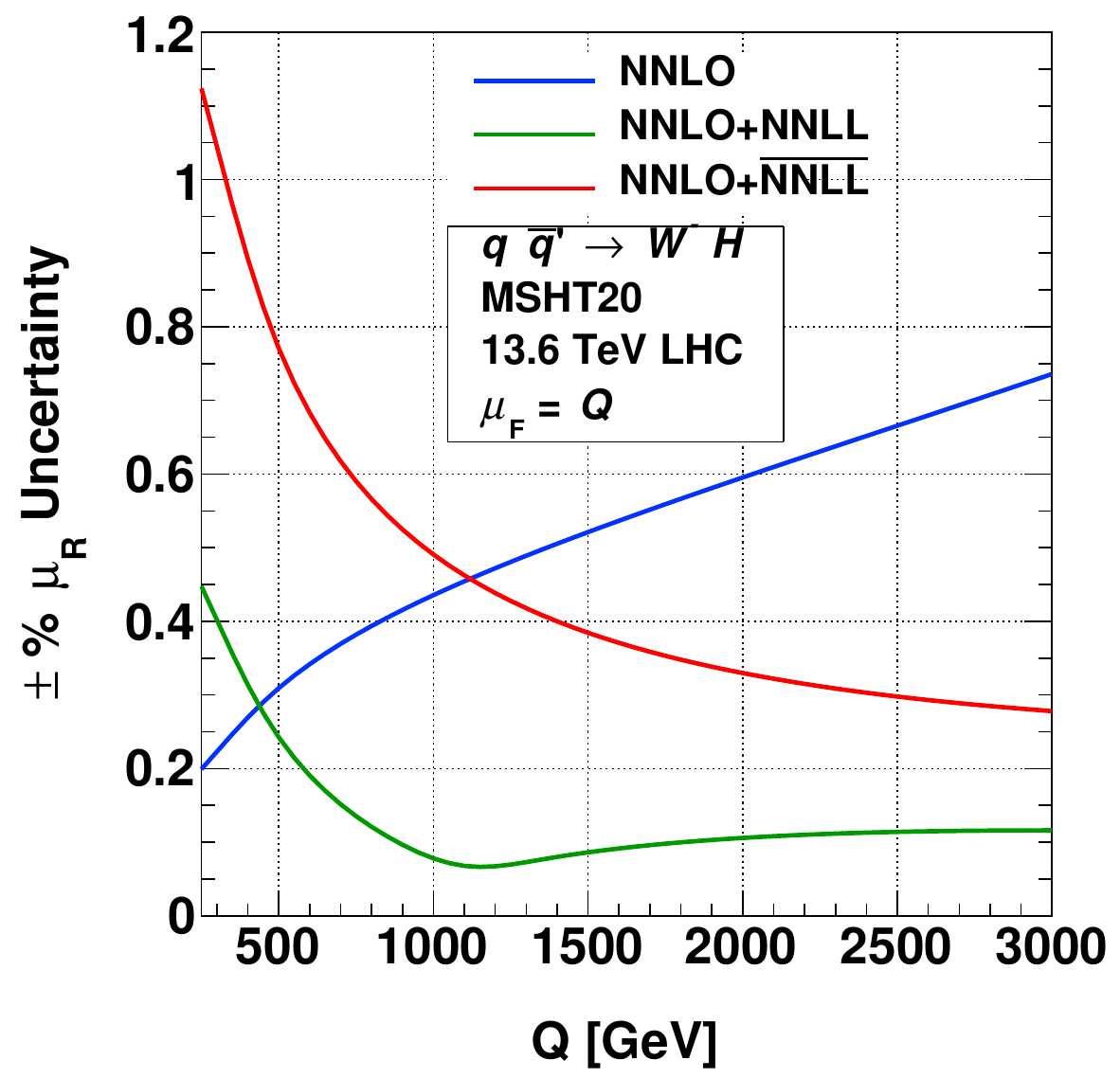}
}
\vspace{-2mm}
\caption{\small{The $\mu_R$ scale uncertainty $ZH$ (left), $W^+H$ (middle) and $W^-H$ (right) are shown at $13.6$ TeV LHC. }}
\label{fig:muR_VH}
\end{figure}

\begin{figure}[ht!]
\centerline{
\includegraphics[scale=0.25]{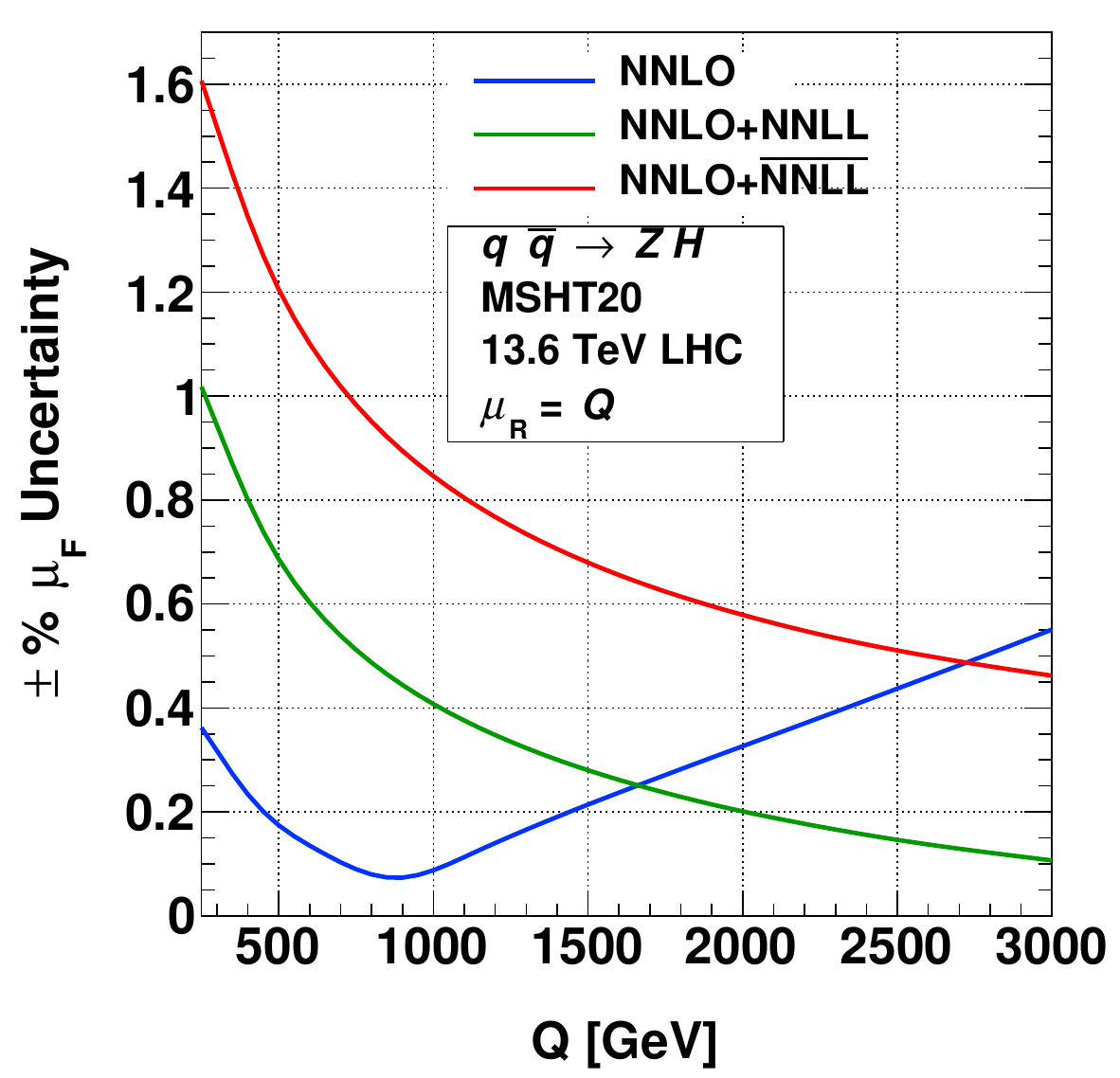}
\includegraphics[scale=0.25]{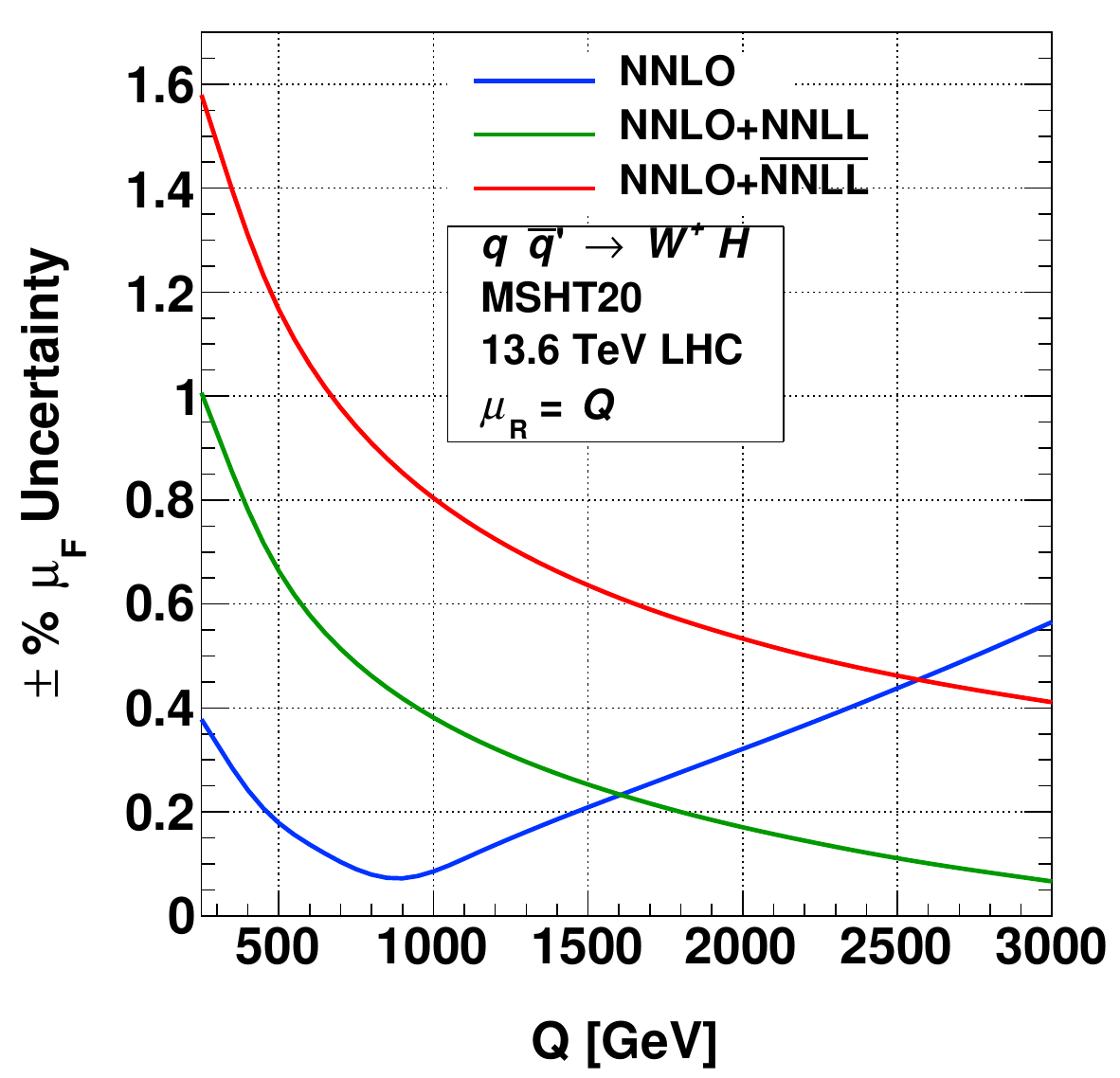}
\includegraphics[scale=0.25]{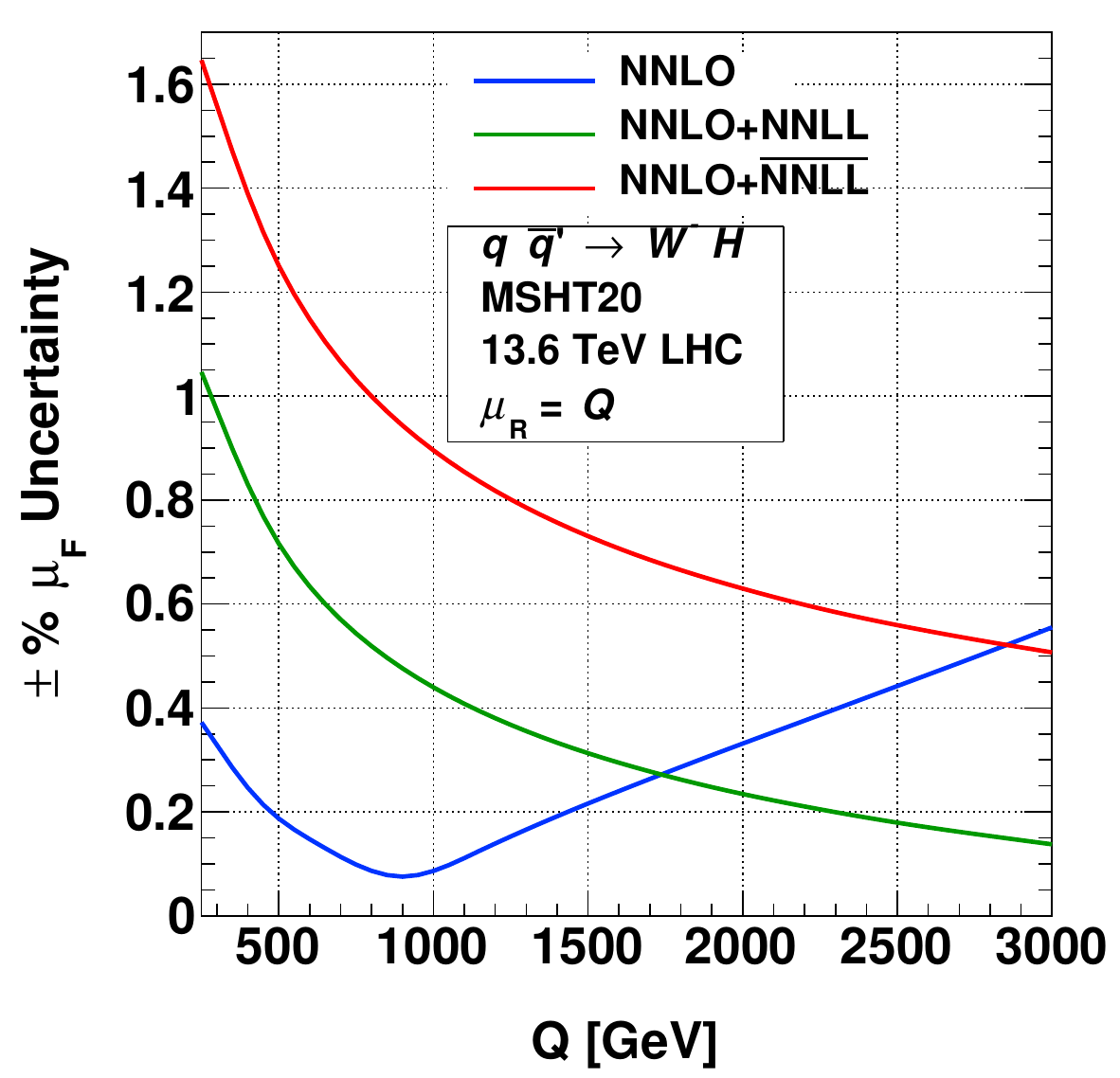}
}
\vspace{-2mm}
\caption{\small{The $\mu_F$ scale uncertainty $ZH$ (left), $W^+H$ (middle) and $W^-H$ (right) are shown at $13.6$ TeV LHC. }}
\label{fig:muF_VH}
\end{figure}

\begin{figure}[ht!]
\centerline{
\includegraphics[scale=0.25]{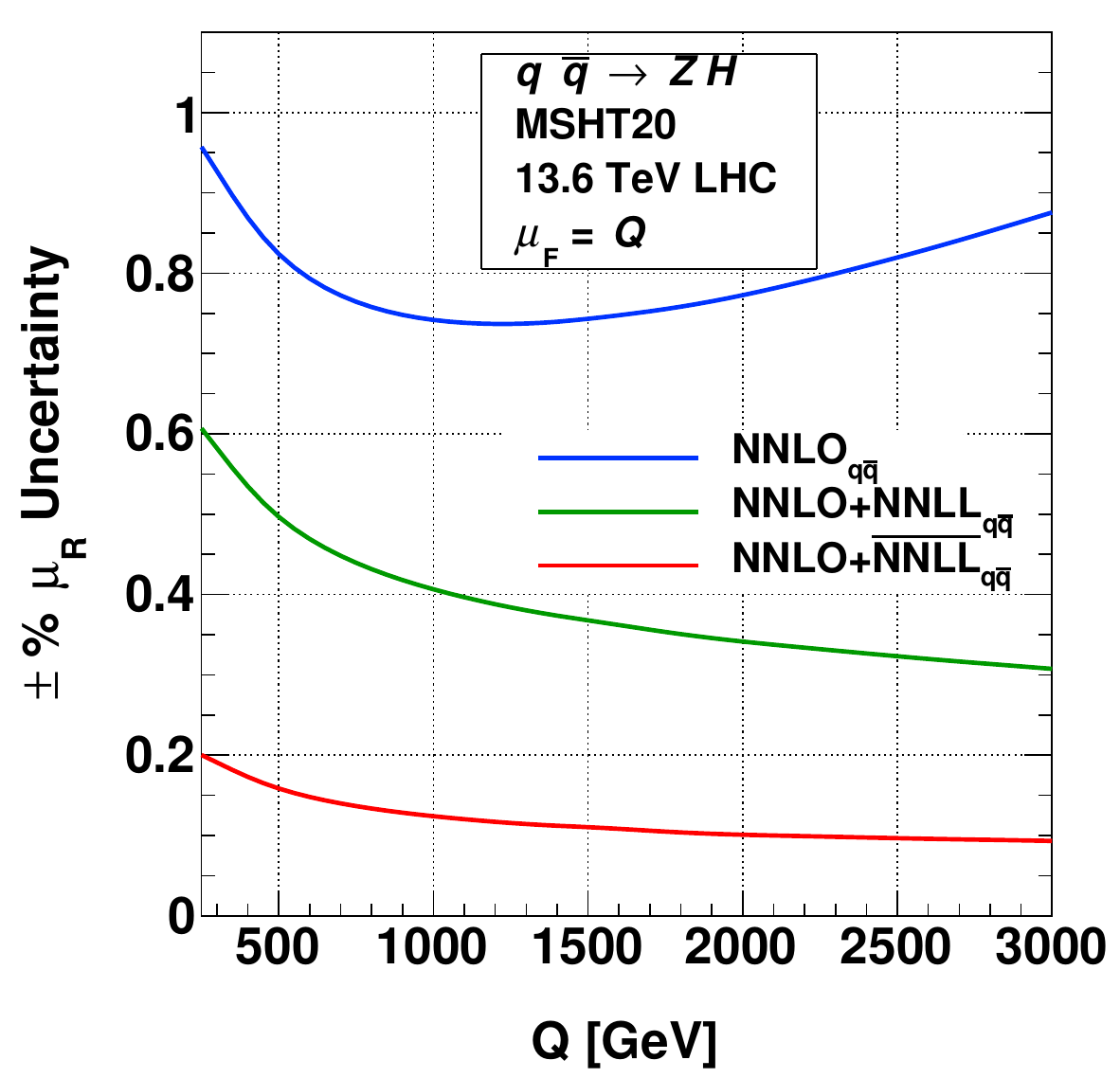}
}
\vspace{-2mm}
\caption{\small{The $\mu_R$ scale uncertainty for FO, SV and NSV are shown for only $q\bar{q} \to ZH$ process at $13.6$ TeV LHC. }}
\label{fig:muR_qqbZH}
\end{figure}

%Since this is observed at NNLO accuracy, the increase in uncertainty can be attributed to the impact of \(qg\) type channel, which starts to affect the results at two-loop accuracy. Since we do not include these effects in the current work, the NSV logarithms associated with these partonic channels accumulate considerable contributions.
%
For completeness, we study the scale variations due to $\mu_R$ and $\mu_F$ separately by varying one and keeping the other fixed at $Q$. Fig.\ \ref{fig:muR_VH}  and fig.\ \ref{fig:muF_VH} depicts the renormalisation scale ($\mu_R$) and factorization scale ($\mu_F$) uncertainties, respectively. These results follow a similar behavior as the $7$-point scale uncertainty results.
To gain a better understanding, we analyse the results from the $q \bar{q}$ channel alone at NNLO, NNLO+NNLL, and NNLO+$\overline{\text{NNLL}}$, as shown in fig.\ \ref{fig:muR_qqbZH}, focusing on the variation of the $\mu_R$ scale. We observe a significant reduction in the percentage of uncertainty as we move from NNLO to NNLO+NNLL and then to NNLO+$\overline{\text{NNLL}}$. Specifically, the percentage of uncertainty decreases from $0.25$ to $0.10$ in the high $Q$ region, as expected.
%This outcome is consistent with the expected behaviour.
%
%
%%%%%%%%%%% Fig: qq -> VH uncertainties
%%%%%%%%%%%%
%%%%%%%%%%% Fig: qq -> VH muR uncertainties
%%%%%%%%%%%%
%%%%%%%%%%% Fig: qq -> VH muF uncertainties
%%%%%%%%%%%%
%%%%%%%%%%% Fig: qq -> ZH (only qqb) muR uncertainties

\begin{figure}[ht!]
\centerline{
\includegraphics[scale=0.25]{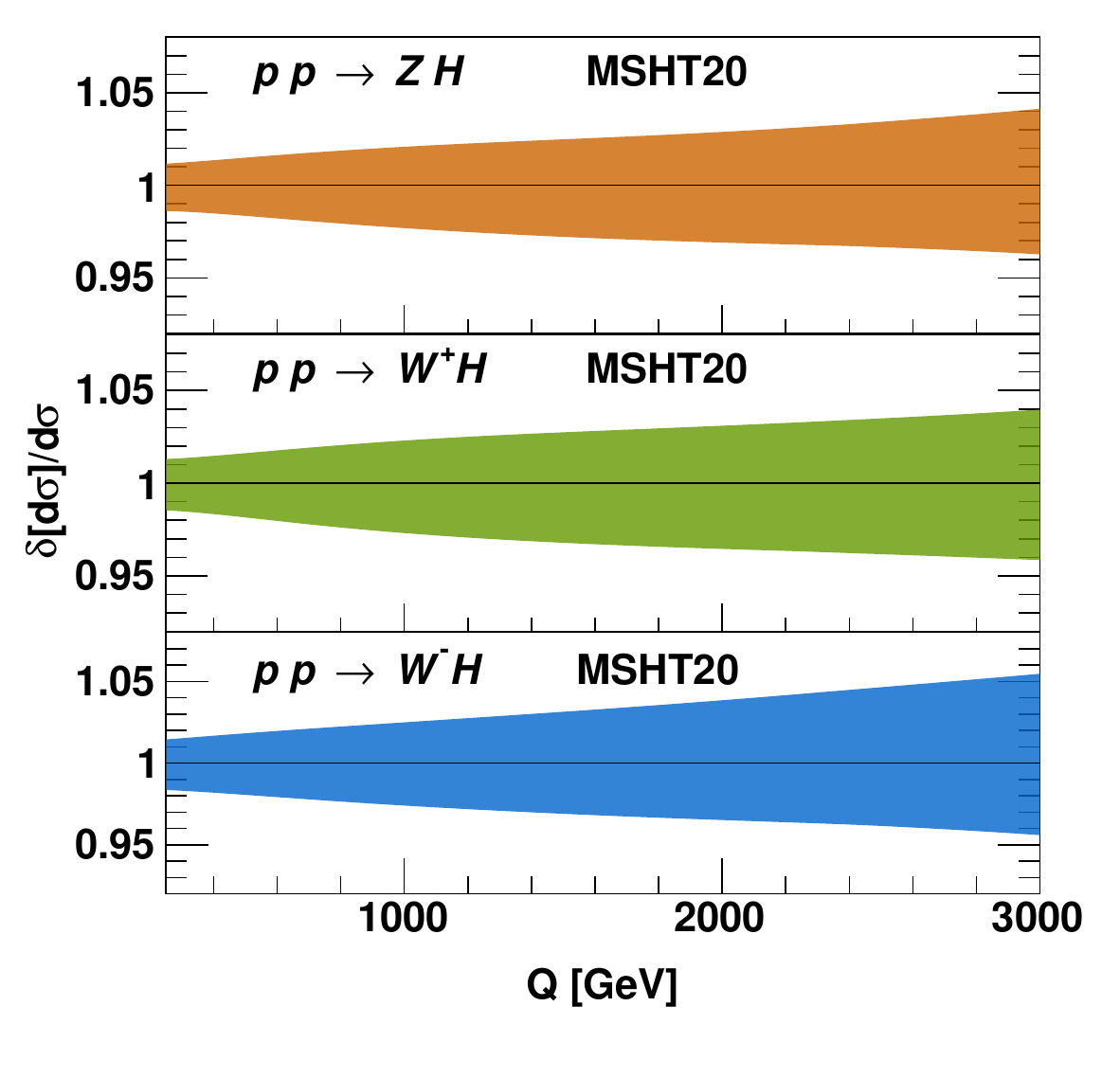}
}
\vspace{-2mm}
\caption{\small{The PDF uncertainty are shown for the $p p \to V H$ process at NNLO+$\overline{\rm NNLL}$ for $13.6$ TeV LHC. }}
\label{fig:PDF}
\end{figure}

We also study the intrinsic PDF uncertainties by calculating the NNLO$+\overline{\text{NNLL}}$ cross section using $64$ different sets of MSHT20 PDFs. For this analysis, we utilise LHAPDF routines. The results are illustrated in fig.\ \ref{fig:PDF}, showing that the uncertainties can reach up to $5\%$ in the $3$ TeV range.
%
%
%%%%%%%%%%% Fig: qq -> VH PDF uncertainties

%%%%%%%%%%%%

%
%
%
%**********************************************************
%             ZH Production
%**********************************************************
% \floatsetup[table]{font=tiny}
\begin{table}[h!]
	\begin{center}{
	\scalebox{1.2}{\begin{tabular}{|c|c|c|c|}
						\hline
						Order&
						$13$ TeV  &  $13.6$ TeV & $100$ TeV \\
						\hline
						\hline
						$\sigma^{\text {NNLO}}_{tot,ZH}$
						&  $0.8566 \pm 1.64\%$  &  $0.9130 \pm 1.66\%$ & $ 11.3200\pm 3.64 \%$ \\
						\hline
						$\sigma^{\text {NNLO + NNLL}}_{tot,ZH}$
						&  $ 0.8597 \pm 0.91\%$ & $0.9163 \pm 0.94\%$ & $11.3475 \pm 2.41 \%$\\
						\hline
						$\sigma^{\text {NNLO}+\overline{\text {NNLL}}}_{tot,ZH}$
						&  $0.8644 \pm 2.17\%  $  & $0.9212 \pm 2.12\%$  & $11.3923 \pm 1.61 \%$\\
						\hline
					\end{tabular}
				}}
				\caption{\small{The $ZH$ production cross sections (in pb) are presented along with corresponding FO, SV and NSV resummed results at different center of mass energy with 7-point scale uncertainties.}
				}
				\label{tab:tableZH}
		\end{center}
	\end{table}
%
%**********************************************************
%             WpH Production
%**********************************************************
% \floatsetup[table]{font=tiny}
\begin{table}[h!]
	\begin{center}{
	\scalebox{1.2}{\begin{tabular}{|c|c|c|c|}
						\hline
						Order&
						$13$ TeV  &  $13.6$ TeV & $100$ TeV \\
						\hline
						\hline
						$\sigma^{\text {NNLO}}_{tot,W^+H}$
						&  $0.9031 \pm 0.43\%$  &  $0.9554 \pm 0.43\%$ & $ 8.8276\pm 1.11 \%$ \\
						\hline
						$\sigma^{\text {NNLO + NNLL}}_{tot,W^+H}$
						&  $ 0.9063 \pm 1.23\%$ & $0.9588 \pm 1.23\%$ & $8.8525 \pm 1.72\%$\\
						\hline
						$\sigma^{\text {NNLO}+\overline{\text {NNLL}}}_{tot,W^+H}$
						&  $0.9112 \pm 3.12\%  $  & $0.9640 \pm 3.12\%$  & $8.8917 \pm 2.89\%$\\
						\hline
					\end{tabular}
				}}
				\caption{\small{The $W^+H$ production cross sections (in pb) are presented along with corresponding FO, SV and NSV resummed results at different center of mass energy with 7-point scale uncertainties.}
				}
				\label{tab:tableWpH}
		\end{center}
	\end{table}

%**********************************************************
%             WmH Production
%**********************************************************
% \floatsetup[table]{font=tiny}
\begin{table}[h!]
	\begin{center}{
	\scalebox{1.2}{\begin{tabular}{|c|c|c|c|}
						\hline
						Order&
						$13$ TeV  &  $13.6$ TeV & $100$ TeV \\
						\hline
						\hline
						$\sigma^{\text {NNLO}}_{tot,W^-H}$
						&  $0.5686 \pm 0.48\%$  &  $0.6063 \pm 0.48\%$ & $ 7.0340\pm 1.10 \%$ \\
						\hline
						$\sigma^{\text {NNLO + NNLL}}_{tot,W^-H}$
						&  $ 0.5710 \pm 1.25\%$ & $0.6088 \pm 1.25\%$ & $7.0552 \pm 1.41 \%$\\
						\hline
						$\sigma^{\text {NNLO}+\overline{\text {NNLL}}}_{tot,W^-H}$
						&  $0.5744 \pm 3.32\%  $  & $0.6125 \pm 3.32\%$  & $7.0899 \pm 3.01\%$\\
						\hline
					\end{tabular}
				}}
				\caption{\small{The $W^-H$ production cross sections (in pb) are presented along with corresponding FO, SV and NSV resummed results at different center of mass energy with 7-point scale uncertainties.}
				}
				\label{tab:tableWmH}
		\end{center}
	\end{table}
    
Finally, in tables \ref{tab:tableZH}, \ref{tab:tableWpH}, and \ref{tab:tableWmH}, we present the total production cross sections for the $VH$ production processes at NNLO, NNLO$+$NNLL, and NNLO$+\overline{\text{NNLL}}$ for center-of-mass energies of $13$ TeV, $13.6$ TeV and $100$ TeV.
In the present context, we consider all these three contributions to the $ZH$ production process and define the total production cross section as
\begin{eqnarray}
        \sigma_{tot,ZH}^{\text{NNLO}} = \text{\ncfodyzh} + \sigma^{gg} (\as^2) + \sigma^{\text{top}} (\as^2) + \sigma^{b\bar{b}} \, 
        \label{eq:ZHtotal_fo}
\end{eqnarray}
\begin{eqnarray}
        \sigma_{tot,ZH}^{\text{NNLO+NNLL}} = \text{\nclldyzh} + \sigma^{gg} (\as^2) + \sigma^{\text{top}} (\as^2) + \sigma^{b\bar{b}} \,
        \label{eq:ZHtotal_resum}
\end{eqnarray}
\begin{eqnarray}
        \sigma_{tot,ZH}^{\text{NNLO}+\overline{\rm NNLL}} = \text{\ncllbdyzh} + \sigma^{gg} (\as^2) + \sigma^{\text{top}} (\as^2) + \sigma^{b\bar{b}} \,
        \label{eq:ZHtotal_resum}
\end{eqnarray}
For the case of $WH$ production process, we define the total production cross section as 
\begin{eqnarray}
\sigma_{tot,WH}^{\text{NNLO}} = \text{\ncfodywh} + \sigma^{\text{top}} (\as^2) \,
        \label{eq:WHtotal_fo}
\end{eqnarray}
\begin{eqnarray}
\sigma_{tot,WH}^{\text{NNLO+NNLL}} = \text{\nclldywh} + \sigma^{\text{top}} (\as^2) \cdot
        \label{eq:WHtotal_resum}
\end{eqnarray}
\begin{eqnarray}
\sigma_{tot,WH}^{\text{NNLO+$\overline{\rm NNLL}$}} = \text{\ncllbdywh} + \sigma^{\text{top}} (\as^2) \cdot
        \label{eq:WHtotal_resum}
\end{eqnarray}
%
%In eqns.\ (\ref{eq:ZHtotal_fo}-\ref{eq:WHtotal_resum}), $\sigma(\alpha_s^i)$ denotes the order of the strong coupling constant $(\alpha_s)$ that the specific channel $(gg,~\text{top},~b\bar{b})$ contributes to the total cross section. 
%
At $13.6$ TeV, the SV resummed results enhance the NNLO ZH production cross section by $0.37$\%, while the NSV results lead to an additional enhancement of $0.53$\%. For $W^{+}H$ production, the enhancement is $0.36$\% from NNLO to NNLO+NNLL and $0.54$\% from NNLO+NNLL to NNLO$+\overline{\text{NNLL}}$. In the case of $W^{-}H$, the SV resummed results show a $0.40$\% enhancement over the NNLO value, and the NSV resummed results provide a further $0.61$\% increase compared to the SV results. Consequently, the inclusion of NSV logarithms results in corrections of approximately $0.89$\% for ZH production, 
$0.90$\% for W$^{+}$H production, and $1.02$\% for W$^{-}$H production compared to the NNLO results at a center-of-mass energy of $13.6$ TeV at the LHC.
We also observe that the cross sections in general increase with increases in the center of mass energy due to the enhancement of available parton fluxes.

%\clearpage
\section{Conclusions}
\label{sec:conclusions}

% \vp{The $VH (V= Z, W\pm) $ production processes has an important place in the Higgs programme at the Large Hadron Collider. We study the Next-to-soft threshold resummation for the $VH$ production as the threshold logarithms not only dominate but also spoil the reliability of the perturbation series in certain kinematical regions. The resummation has been performed up to the $\overline{\text{NNLL}}$ accuracy by matching with the fixed order NNLO results (obtained from \texttt{vh@nnlo}). We present phenomenological results for the invariant mass distribution (13.6 TeV) and the total production cross section (13,13.6 and 100 TeV ) up to the NNLO$+\overline{\text{NNLL}}$ accuracy.} 

In this work, we have investigated the impact of next-to-soft (NSV) threshold resummation on associated Higgs production with a vector boson ($VH$, where $V = Z, W^\pm$), a crucial process in the Higgs precision program at the LHC. By performing resummation at $\overline{\text{NNLL}}$ accuracy and matching our results with FO NNLO calculations, we provide precise predictions for the invariant mass distribution at 13.6 TeV and total production cross sections at $13$, $13.6$, and $100$ TeV. 

The precise computation of threshold corrections continues to be an area of significant theoretical interest, particularly in understanding the role of NSV effects and their factorisation properties \cite{Vernazza:2023hrf,Broggio:2023pbu,vanBeekveld:2021hhv,Bahjat-Abbas:2019fqa}. In this study, we employ the generic resummation approach developed in \cite{AH:2022lpp,AH:2020iki} to systematically compute both SV and NSV resummed corrections to $q \bar{q} \rightarrow VH$ production at two-loop accuracy. The foundation of our work lies in the observation that for diagonal channels, the SV+NSV resummed structure depends solely on the initial parton channels, and the process-dependent information is encapsulated in the coefficient $C_0$ of eqn. \ref{DeltaN}. This framework validated up to NNLO for Higgs and DY production \cite{AH:2020iki,AH:2022lpp,AH:2020qoa,Ahmed:2020nci,AH:2021kvg,AH:2021vdc,AH:2021vhf} is applied here to a $2 \rightarrow 2$ process of form factor-type diagrams, demonstrating results that align with theoretical expectations.
We observe that in the high-Q region, the NSV resummation contributes an additional $2\%$ of the LO to the NNLO results and further, it reduces the $\mu_R$ scale uncertainty from $0.25\%$ to $0.10\%$ in the high $Q$ region for the $q\bar{q}$-channel. 
Given the increasing demand for high-precision theoretical predictions in upcoming LHC runs and future collider experiments, the results provided in this work 
pave the way for their broader application in Higgs phenomenology and beyond.
% our findings reinforce the predictive power of resummation techniques for precision collider physics, paving the way for their broader application in Higgs phenomenology and beyond.
These refined predictions significantly enhance the reliability of theoretical calculations, facilitating more accurate extractions of Higgs couplings from experimental data.
These developments will be instrumental in probing potential deviations from the SM and exploring new physics scenarios with greater sensitivity.
%Our work provides a foundation for extending resummation formalisms to even higher accuracy.

% \appendix
% \section{Appendix}
% If needed, add appendices here.
% \ab{Suggestion to add Analytical expressions for 
%     \begin{itemize}
%         \item $\Delta_{cc,N}$
%         \item $C_0$ 
%         \item $\Psi_{SV,N}$
%         \item $\Psi_{NSV,N}$
%     \end{itemize}}

% Acknowledgments
\begin{acknowledgement}
%    Acknowledge funding sources, collaborations, or other contributions here.
A.B.\ acknowledge the financial support by the Generalitat Valenciana, the spanish goverment and ERDF funds from the European Commission (CNS2022‐136165, funded by MCIN/AEI/10.13039/501100011033/\ and by the European Union “NextGenerationEU/PRTR”). 
The research work of M.C.K.\ is supported by the SERB Core Research Grant (CRG) under the project CRG/2021/005270. 
\end{acknowledgement}

% Bibliography
\bibliographystyle{unsrt}
\bibliography{references} % Provide a .bib file with your references

\end{document}